\documentclass[conference]{IEEEtran}
\ifCLASSINFOpdf
\else
\fi
\usepackage{amsmath}
\usepackage{mathptmx}
\usepackage{graphicx}
\usepackage[tight,footnotesize]{subfigure}

\usepackage{algorithm}
\usepackage{algpseudocode}
\newcommand{\argmin}[1]{\underset{#1}{\operatorname{argmin}}\;}
\newcommand{\remove}[1]{}

\usepackage{graphicx,color}
 
\newcommand{\dd}[1]{}

\begin{document}
%
\title{On Bootstrapping Machine Learning Performance Predictors via Analytical Models}

\author{\IEEEauthorblockN{Diego Didona and Paolo Romano}
\IEEEauthorblockA{INESC-ID / Instituto Superior T\'{e}cnico, Universidade de Lisboa}}


%


\maketitle

\begin{abstract}
Performance modeling typically relies on two antithetic methodologies: white box models, which exploit knowledge on system's internals and capture its dynamics using analytical approaches, and black box techniques, which infer relations among the input and output variables of a system  based on the evidences gathered during an initial training phase.
In this paper we investigate a technique, which we name {\em Bootstrapping}, which aims at reconciling these two methodologies and at compensating the cons of the one with the pros of the other. We thoroughly analyze the design space of this gray box modeling technique, and identify a number of algorithmic and parametric trade-offs which we evaluate via two realistic case studies, a  Key-Value Store and a Total Order Broadcast service.  
\end{abstract}


%
\IEEEpeerreviewmaketitle

\section{Introduction}

In the era of cloud computing, performance modeling of distributed systems plays a role of paramount importance. Not only  does it serve for traditional purposes, such as capacity planning~\cite{Menasce:CP} and anomaly detection~\cite{pipL}. 
By allowing the definition of self-tuning and automatic resource provisioning schemes, performance forecasting tools represent also a fundamental building block of the elastic computing paradigm.


Classical approaches to performance prediction rely on two, antithetic, techniques: Machine Learning (ML)~\cite{Ahmad11L,Duggan14L,Couceiro10} and Analytical Modeling (AM)~\cite{Urgaonkar:2005,Singh:predico,tay:book}.

ML-based techniques embody the \emph{black box} approach, which infers performance models based on the relations among the input and output variables of a system  that are observed during an initial training phase. ML-based performance models can typically achieve a very good accuracy when working in interpolation, i.e., in areas of the features' space that have been sufficiently explored. On the downside, the accuracy of such techniques is typically hindered when used in extrapolation, i.e., to predict values in regions of the parameters' space not observed during the training phase. Another major issue of ML techniques is that the number of configurations to be explored grows exponentially with the number of variables (often referred to as features, in the ML literature) characterizing the application --- the so-called \emph{curse of dimensionality}~\cite{Bishop:ML}. This has a direct impact on the time needed to gather a sufficiently representative training set, which can quickly become large enough to make the usage of such techniques cumbersome or even prohibitive in complex systems.

Analytical models, conversely, are based on \emph{white box} approaches, according to which the model designer exploits knowledge about the dynamics of the target system in order to mathematically express its input/output relations. Analytical models require no or minimal training phase. On the other hand, in order to allow for mathematical tractability, they rely on approximations and simplifying assumptions. Hence, the accuracy of analytical models can be challenged in scenarios in which such approximations and assumptions are not valid.

Being based on radically different techniques, AM and ML have been seen for decades as competitive approaches to perform performance forecasting. Over the last years, however, we have witnessed an increasing number of proposals based on \emph{gray box} approaches, aimed at reconciling these two paradigms. The ultimate goal of these techniques is to achieve the best of the two worlds, namely the extrapolation capabilities of AM, combined with the high accuracy of ML when working in interpolation (i.e., once sufficient information on actual system's performance has been gathered).

In this paper we investigate a technique, which we name the \emph{Bootstrapping}, whose key idea consists in relying on an analytical model to generate a synthetic training set over which a complementary machine learner is initially trained. The synthetic training set is then updated over time to incorporate new samples collected from the operational system. By exploiting the knowledge of the white box analytical model, the resulting model inherits its initial prediction capabilities, avoiding, unlike traditional ML-based approaches, the need for any preliminary observation of the system in operation prior to their instantiation. At the same time, by updating the synthetic knowledge base with samples coming from the actual system, the bootstrapping technique allows for progressively correcting initial errors due to inaccuracies of the analytical model. Furthermore, the white box analytical model allows for enhancing the robustness of the resulting gray box predictor, by improving its accuracy in regions of the features' space not observed during the training phase.

The idea at the basis of Bootstrapping  has been used in several recent works~\cite{ironmodel,Romano:batchingL,Rughetti:ccgridL,Tesauro:hybridL} in the area of performance modeling of complex systems, which have highlighted the potentiality and relevance of this technique. However, the design space of the Bootstrapping approach includes a number  
of algorithmic and parametric trade-offs, which can have a strong impact on both accuracy and construction time of the resulting gray box model, and which were never identified or discussed in the literature.

In this paper we fill this gap by presenting what is, to the best of our knowledge, the first detailed algorithmic formalization of this technique. We identify two key choices in the design of bootstrapping algorithms:\vspace*{.2cm}\\
\noindent$i)$ how many samples of the output of the analytical model should be used to populate the initial synthetic training set;\vspace*{.1cm}\\
\noindent$ii)$ which algorithmic techniques should be used to update the (initially fully) synthetic knowledge base with new evidences gathered from the operational system.

We propose a set of alternative approaches to tackling these two issues, and evaluate the impact of these alternatives by means of an extensive experimental study based on two case studies: a popular distributed Key-Value Store (Infinispan by Red Hat~\cite{infinispan}) and a Total Order Broadcast (TOB) service~\cite{distributedSystemBookL}. The former is representative of typical cloud data stores, whose performance exhibits complex non-linear trends and is affected by a large number of parameters. The latter represents an incarnation of the consensus problem~\cite{paxos} and is used as a fundamental building block in a number of fault-tolerant approaches~\cite{PedoneStateMachineL,d2stmL}.
We consider two recent  analytical models for these systems~\cite{Didona:prompt, Romano:batchingL}, which we instantiate using different parametrizations, hence emulating scenarios in which the white-box models achieve different degrees of accuracy (e.g., due to noisy measurements during the white-box model initialization phase).

Our experimental results confirm the actual potentiality of this technique, but also shed light on several pitfalls and on the relevance of correctly tuning  a number of parameters: these are issues that,  to the best of our knowledge, were never discussed in the literature and for which we propose and evaluate  alternative solutions. 

The remainder of this paper is structured as follows. Section~\ref{sec:relatedWork} discusses related work. We provide the algorithmic formalization of Bootstrapping in Section~\ref{sec:bootstrapping}. Section~\ref{sec:expEval} is devoted to presenting the case studies and the experimental evaluation. Finally, Section~\ref{sec:conclusions} concludes the paper.

\section{Related work}
\label{sec:relatedWork}
Different approaches have been proposed, in the literature, that leverage on AM and ML in synergy. These approaches differ in the way they combine AM and ML, as well as for the employed learning methodology -- e.g., off-line vs on-line learning (based, for example, on Reinforcement Learning, RL) 
-- and algorithm -- e.g., Artificial Neural Networks (ANNs) vs Decision Trees (DTs) vs Support Vector Machines (SVMs).

\begin{figure}[t]
\centerline{
\subfigure[Initialization phase.]{\label{fig:intro1}\includegraphics[scale=.19]{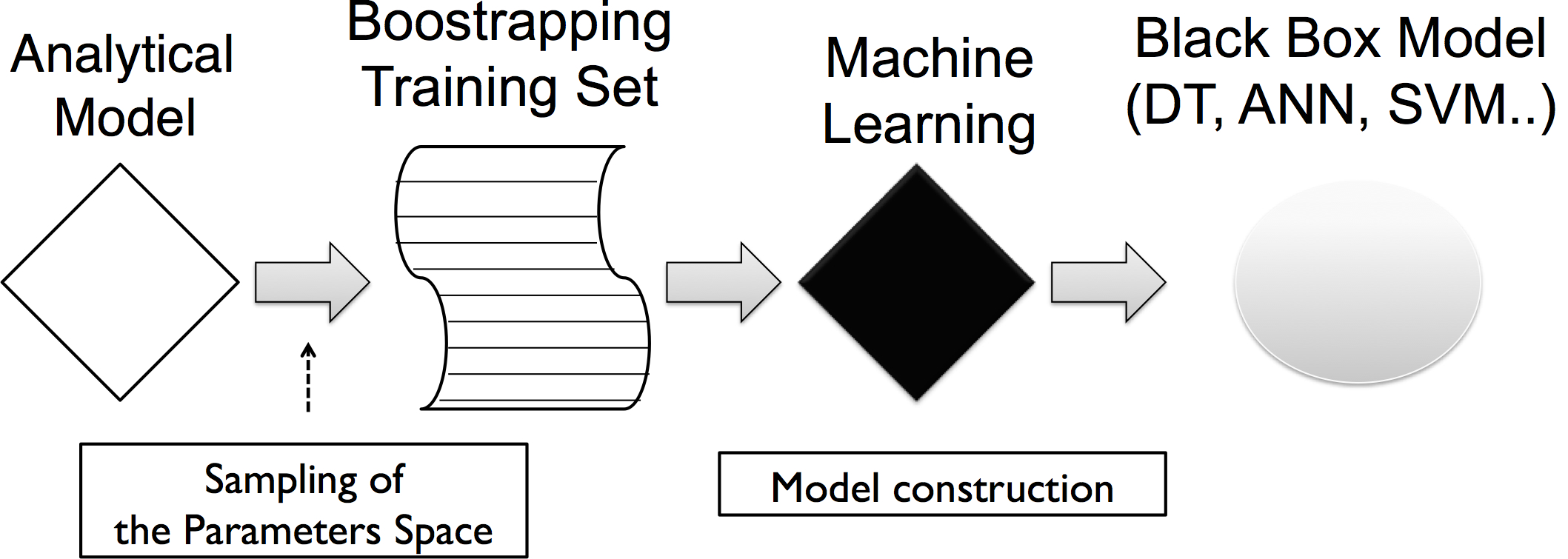}}
\hfil
\hspace{5mm}
\subfigure[Updating]{\label{fig:intro2}\includegraphics[scale=.19]{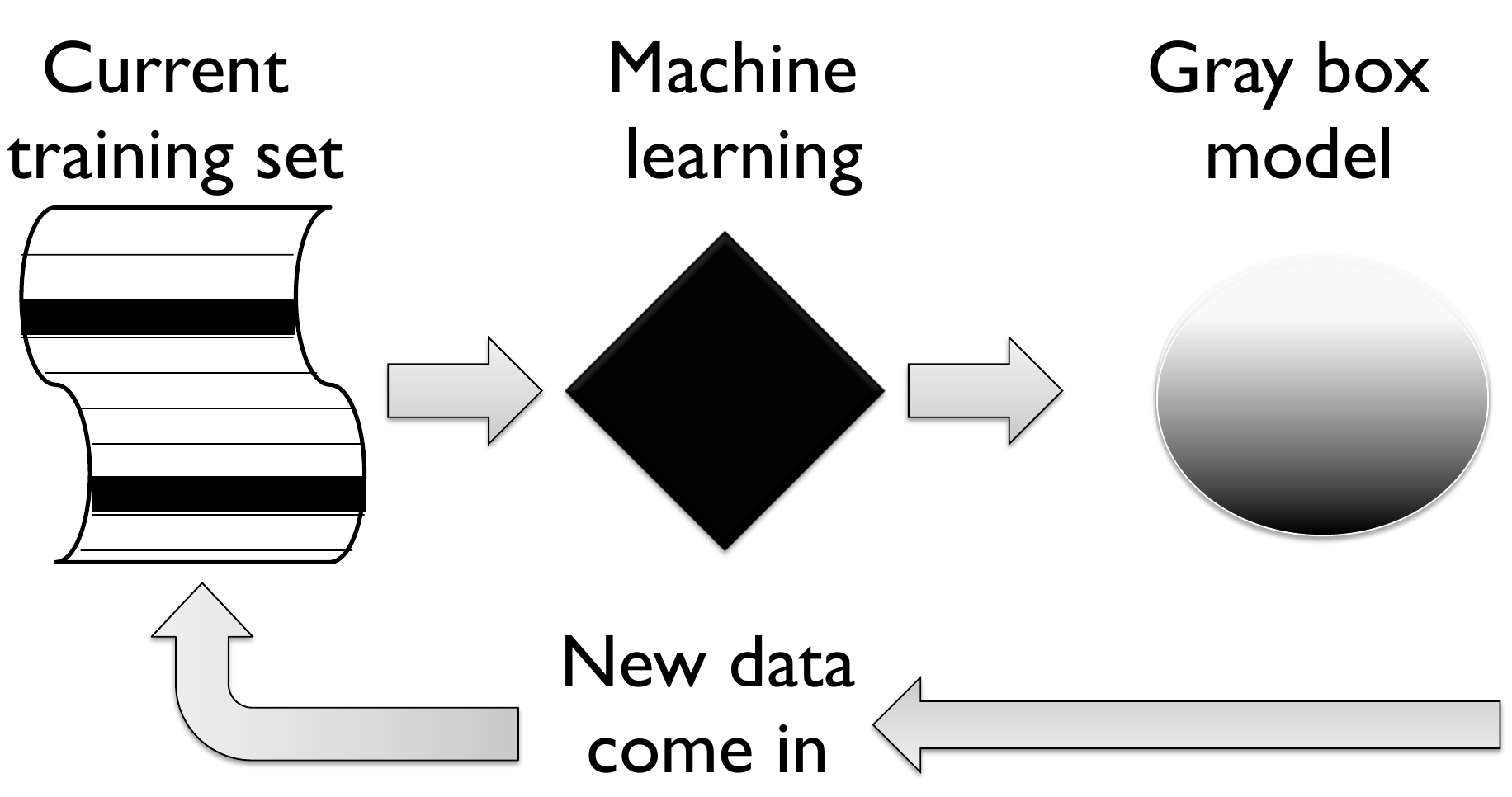}}
   }
   \caption{Main phases of the Bootstrapping technique}
   \label{fig:intro}
\end{figure}

The technique that we investigate in this paper, and which we call Bootstrapping, is one of such approaches, and variants of this idea have been already applied with success to a few case studies in the area of performance modeling of complex systems. For instance, in the work by Tesauro et al.~\cite{Tesauro:hybridL} the problem of provisioning a platform in order to meet a target Quality of Service is cast to a Markov Decision Problem that is solved by the means of RL. The inner states of the learner are initialized according to the output of a closed or open network of queues. Romano and Leonetti~\cite{Romano:batchingL} apply the idea of bootstrapping the knowledge base of RL algorithm to automate the tuning of the batching level of a Sequencer-based Total-Order protocol. The system is first modeled as a $M/M/1$ open queue; multiple instances of the UCB~\cite{Auer:UCBL} RL algorithm are, then, employed at runtime to refine the model.  Schroeder et al.~\cite{Schroeder:MPLL} model a database as a $M/H_2/1$ queue in order to determine an initial value of the Multiprogramming Level (MPL), which is then refined on-line by means of a hill climbing algorithm. In a recent work by Rughetti et al.~\cite{Rughetti:ccgridL}, the bootstrapping methodology is employed in order to predict the response time of Transactional Memory-based applications depending on the number of running threads. The analytical model relies on a set of functions whose parameters are fitted depending on the samples gathered at runtime; the employed machine learner is a backward propagation ANN.

The bootstrapping technique has also been employed to detect software runtime misbehaviors: in IronModel~\cite{ironmodel}, a Queueing Theory-based model is used to bootstrap the knowledge base of a DT regressor to predict the response time of various components in a data centre. Upon detecting a deviation of the measured latencies with respect to the predicted ones for a component under a certain workload, the system administrator checks whether there is a bug in the software of the component. If this is not the case, the relevant traces are fed to the DT; the machine learner is, then, able to generate a new rule to incorporate the new knowledge, by splitting a leave on the tree depending on the feature that is found to be more correlated to the mis-prediction.

With respect to these papers, which present examples of exploitation of the Bootstrapping method, this work is the first to provide a rigorous algorithmic formalization of this technique, and to explore, in a systematic fashion, a number of complex trade-offs in its design space. Our experimental evaluation allows us to gain insights on the sensitivity of the Bootstrapping technique to the configuration of internal parameters and to alternative algorithmic variants.

This work is clearly related also to other modeling techniques, different from the Bootstrapping one, that rely on a combination of white and black box models. For instance, Zhang et al.~\cite{Zhang:Regression}, starting from the Utilization Law~\cite{kleinrock}, exploit regression to estimate jobs' resource demands in multi-tier systems in order to instantiate a queuing network model. TAS~\cite{Didona:taas,Didona:prompt} is a system for predicting performance of distributed in-memory data stores that leverages on AM and ML by taking a different approach, called {\em divide and conquer} approach: AM is exploited to capture the effect on data and CPU contention, whereas a DT regressor is exploited to predict the latency of network bound operations (e.g., two-phase commit execution time). 
Another class of hybrid solutions to the performance prediction problem relies on combining white and black box models into ensembles. A first approach of this kind consists into exploiting cross-validation or a classifier to identify which is the best predictor to use depending on the incoming query~\cite{icpe15,chen13}. A second approach consists into exploiting black-box models to correct the inaccuracies of a base white-box one; this is accomplished by training the black box learners over the residual errors of the white box one, rather than on the target KPI function directly~\cite{icpe15,Didona:computing}.
Finally, the Elastisizer framework \cite{no-one-cluster-sizeL} exploits a DT regressor to predict running time of Map-Reduce jobs in Cloud environments; AM is exploited to compute some metrics that are highly correlated with the target one and that are fed to the DT as additional input features.

\section{The Bootstrapping Technique}
\label{sec:bootstrapping}

In this section we describe the Bootstrapping technique in a top-down fashion: we first overview the overall execution of the algorithm, encapsulating several relevant building blocks into abstract primitives. Next, in Sections~\ref{sec:init} and~\ref{sec:update}, we shall discuss in detail the key parametric and algorithmic trade-offs associated with each of these primitives.

\begin{algorithm}[t]
\scriptsize
  \caption{Bootstrapping main loop}
  \label{alg:main}
  \begin{algorithmic}[1]
  \Function{Main}{\null}
   \State $ML\ ml$\Comment{The machine learner}
   \State $AM\ am$\Comment{The analytical model}
   \State $DataSet\ ST = initKB()$\Comment{Generate the synthetic training set}
   \State $ml.train(ST)$\Comment{Train the ML over the synthetic training set}
   \While{\textbf{true}}
   \State $DataSet\ D = collectSamples()$\Comment{Collect samples at runtime}
   \State $updateKB(ST,D)$\Comment{Incorporate the new samples in the knowledge base}
   \State $ml.train(ST)$\Comment{Re-train the ML over the updated training set}
   \EndWhile 
   \EndFunction
   \item[]
   \Function{query}{Configuration x}
   \State \textbf{return} $ml.query(x)$
   \EndFunction   
  \end{algorithmic}
  \end{algorithm}
As reported in the pseudo-code Alg.~\ref{alg:main}, the Bootstrapping technique consists of two main phases: the initialization of the black box model based on the predictions of the analytical one (lines 4-5), and its re-training, which is performed every time that new samples from the running application (lines 6-10) become available, and which incorporates them into the knowledge base (lines 6-10).

The initialization phase, depicted in Fig.~\ref{fig:intro1} and detailed in Sec.~\ref{sec:init}, is composed, in its turn, of three steps:\vspace*{.1cm}\\
\noindent $i)$ \textit{Sampling of the parameters' space of the AM:} first of all we need to determine a subset $T$ of the parameters' space of the AM, which is used to bootstrap the knowledge base of a machine-learner. As already mentioned, the number of samples of a multi-dimensional space that are necessary to characterize an arbitrary function defined over this space grows, generally speaking, exponentially with the dimensionality of the space. This step is, thus, aimed at determining \emph{how many} samples to include in the initial synthetic training set in order to have a sufficient coverage of the whole parameters' space. This step will be further detailed in Sec.~\ref{sec:init}. \vspace*{.1cm}\\
\noindent$ii)$ \textit{Generation of the synthetic training set:} the analytical model is queried in order to compute a prediction of the performance of the application for each of the samples in $T$. The output of this phase is a new set $ST$, whose elements are tuples of the form $<x,am.query(x)>$, where $x\in T$ is an element of $T$ and $am.query(x)$ is the corresponding prediction computed by the analytical model.\vspace*{.1cm}\\
\noindent $iii)$ \textit{Black box model construction}: the ML is trained on the dataset $ST$ and produces a statistical model of the application's performance. It should be noted that the Bootstrapping technique can be used in conjunction with alternative ML techniques, such as DTs, ANNs, etc. 

The update phase, illustrated in Fig.~\ref{fig:intro2} and detailed in Sec.~\ref{sec:update}, consists of two steps: \vspace*{.1cm}\\
\noindent $i)$ \textit{update of the training set}: the $ST$ set is updated in order to incorporate knowledge represented by the samples coming from the running application. There are several ways to perform this operation: Sec.~\ref{sec:update} will be devoted at describing various alternatives; \vspace*{.1cm}\\
\noindent $ii)$ \textit{black box model construction}: the ML is trained on the updated dataset $ST$ and produces a new statistical model of the application.

\subsection{Synthetic Knowledge Base Initialization}
\label{sec:init}

The first step of the Bootstrapping technique is embodied by the {\sc initKB} function, whose pseudo-code is reported in Alg.~\ref{alg:init}. This function performs two main operations. The first one consists in selecting a subset of samples from the whole space of possible configurations for the application. The second one consists in generating the synthetic training set by exploiting the predictions output by the analytical model for each of the elements in this subset.

The sampling operation, executed by the function {\sc sampleConfigSpace} in Alg.~\ref{alg:init}, has to determine \emph{how many} samples to select from the configurations space, such that the resulting synthetic training set (which has to be learnt by a ML-based regressor) is representative of the target performance function to be modeled. The choice of the number of samples to use can affect significantly the effectiveness of the Bootstrapping methodology. A low number of samples allows for reducing the duration of the training phase; also, it may favor the subsequent update phase of the training set: the lower the number of synthetic samples, the higher the relative density of the real samples in the updated training set. This can reduce the time it takes for the real samples to outweigh the synthetic ones, and correct possible errors of the analytical model. However, using a lower number of synthetic samples also yields the black box model to approximate more coarsely the original white box one, which may degrade accuracy. On the other hand, a very large training set provides more detailed information to the black box learner on the function embodied by the analytical model, and can favor a better approximation of such function. However, it comes with the downside of an increased training time and a longer transient phase before runtime samples can take over synthetic ones.

\begin{algorithm}[t!]
\scriptsize
  \caption{Initialization phase}
  \label{alg:init}
  \begin{algorithmic}[1]
     \Function{initKB}{}    
    	\State $Set\ T = SampleConfigSpace()$\Comment{Training configurations}
      	\State $DataSet\ ST=\emptyset$\Comment{AM-based training set}
      	\ForAll{$x \in T$}
      		\State $ST = ST\cup \{x,am.query(x)\}$
      	\EndFor
      	\State \textbf{return} $ST$
    \EndFunction    
\end{algorithmic}
\end{algorithm}

Unlike previous works on Bootstrapping, which do not tackle this issue, we propose a a cross-validation based algorithm that evolves by iteratively performing the following steps. First, a training set $S$ is generated: $N$ samples are drawn uniformly at random from the whole parameters' space and the AM is queried to predict the output corresponding to each of such points. Then, the ML accuracy over $S$ is evaluated via ten-fold cross validation. This entails partitioning $S$ into 10 bins $S_1 \ldots S_{10}$ and then, iteratively for $i=1 \ldots 10$, training the ML over $S \setminus S_i$ and evaluating its accuracy against $S_i$. 
If the average accuracy over the 10 rounds falls beyond a threshold $\epsilon$, the algorithm stops and $S$ becomes the initial synthetic training for the bootstrapped black box learner. Otherwise, a new set $S'>S$ is chosen and another iteration of the algorithm is performed.
\subsection{Update of the Knowledge Base}
\label{sec:update}

The {\sc updateKB} function, reported in Alg.~\ref{alg:update}, is the core of the Bootstrapping methodology, as it allows for the incremental refinement of the initial performance model. This function is responsible for incorporating the knowledge coming from the running application into the initial synthetic training set, by gradually correcting inaccurate performance estimations of the original model. 

The {\sc updateKB} function takes in input the dataset $D$ containing new samples and injects these samples into the current training set. The key issue here is that the new samples contained in $D$ may contradict the synthetic samples generated by the AM, which are already present in the training set --- this is the case when $D$ contains samples belonging to regions of the features' space in which the AM achieves unsatisfactory accuracy. In this work, we consider two complementary techniques that aim at reconciling possible divergences  between synthetic and actual samples: $weighting$ and $replacing$. Weighting is a well-known and widely employed technique in the ML area~\cite{weighting}: the higher the weight for a sample, the more the ML will try to minimize the fitting error around it when building the statistical model. In the Bootstrapping case, weighting can be used as a means to suggest the ML to give more relevance and trust to real samples than to synthetic ones. Another complementary approach consists in removing pre-existing ``close enough'' (synthetic) samples from the training set, whenever we incorporate new observations drawn from the operational system.

To the best of our knowledge, no previous work investigates the effectiveness of weighting in the context of the bootstrapping technique. Moreover, we consider four implementations of the {\sc updateKB} function (three of which are novel) that incorporate new knowledge according to different principles. We describe these techniques in the following.

\begin{algorithm}[t]
\scriptsize
\caption{Update phase}
\label{alg:update}
\begin{algorithmic}[1]
\Function{updateKB}{DataSet D}
    	\State $setWeight(D,w)$\Comment{Set the weight to the new samples}
    	\State {\bf function} update = any function in \{\textsc{Merge,RNN,RNR,RNR2}\}
    	\State $update(D)$;
    	\State $ml.train(ST)$\Comment{Retrain the ML with the new dataset}
    \EndFunction
    \item[]
    
    \Function{merge}{DataSet D}
    	\State $ST = ST \cup D$\Comment{Add the real samples}
    \EndFunction
    \item[]
    
    \Function{RNN}{DataSet D}
    	\ForAll{$(x,y)\in D$}
    		\State $(x_r, y_r) = \argmin{(x',y')\in ST}\{dist(x',x)\}$\Comment{Find the NN}
    		\State $ST = ST\setminus\{(x_r,y_r)\}$\Comment{Remove the NN}
    		\State $ST = ST\cup\{(x,y)\}$\Comment{Insert the real sample}
    	\EndFor
    \EndFunction
    \item[]
    
    \Function{RNR}{DataSet D, double c}
    \State $DataSet\ D\_NR = \emptyset$\Comment{Temporary NN set}
    	\ForAll{$(x,y)\in D$}
    		\State $D\_NR = \{(x_t,y_t)\in ST:dist(x,x_t)\leq c \land isSynthetic(x_t,y_t)\}$
    		\State $ST = ST \setminus D\_NR$\Comment{Remove the NNs}
    		\ForAll{$(x,y)\in D$}
    			\State $ST = ST\cup\{(x,y)\}$\Comment{Add real samples}
    		\EndFor    	
    	\EndFor
    \EndFunction
    \item[]
    
    \Function{RNR2}{DataSet D, double c}
    \State $DataSet\ D' = D$\Comment{Temporary set of real samples still unmatched} 
    	\ForAll{$(x_t,y_t)\in ST$}
    		\State $(x_r, y_r) = \argmin{(x,y)\in D}\{dist(x_t,x)\}$\Comment{Find the NN}
			\If{$dist(x_r,x_t)\leq c \land isSynthetic(x_t,y_t)$}    		
    			\State $ST = ST\setminus\{(x_t,y_t)\}$\Comment{Remove the NN}
    			\State $ST = ST \cup \{(x_t,y_r)\}$\Comment{Add NN with modified output}
    			\State $D' = D'\setminus{(x_t,y_t)}$\Comment{Remove the real sample from the temp.~set}
    		\EndIf
    	\EndFor
    		\ForAll{$(x,y)\in D'$}
    			\State $ST = ST\cup\{(x,y)\}$\Comment{Add unmatched samples}
    		\EndFor   
    \EndFunction
    
    \normalsize
    \end{algorithmic}   
    \end{algorithm}

\noindent
\textbf{Merge.} This is the simplest variant that we consider, and it consists in adding  the new samples to the existing set $ST$ (lines 7-9). This implies the possible co-existence of real and synthetic samples that map very similar input features to very different performance. Hence, the use of weights is the only means to induce the ML to give more importance to real samples over (possibly contradicting) synthetic ones. 
\vspace*{.2cm}\\~
\textbf{Replace based on Nearest Neighbor (RNN).} To the best of our knowledge, this algorithm was first used by Rughetti et al.~\cite{Rughetti:ccgridL}. It consists of two steps, which are repeated for each element $(x,y)$ in $D$: i) find the element $(x_r,y_r)$ that is closest (using the Euclidean distance) to $(x,y)$ in $ST$ (line 12) and ii) replace $(x_r,y_r)$ with $(x,y)$ (lines 13-14). Unlike the original proposal, also in this case we allow the newly injected sample to receive a weight $w$.
 Note that, once an element from $D$ is inserted in $ST$, it becomes eligible to be evicted from the set, even in favor of another sample contained in $D$ itself. This algorithm aims at progressively replacing all the synthetic samples from $ST$ with real ones; by switching a real sample with its nearest neighbor in $ST$, moreover, this algorithm aims at keeping unchanged the density of samples in $ST$. 
\vspace*{.2cm}\\~
\textbf{Replace based on Nearest Region (RNR).} This algorithm represents a variant of RNN. A first difference is that, in order to avoid ``losing" knowledge gathered from the running system, RNR policy only evicts synthetic samples from the training set. Moreover, instead of replacing a single sample in $ST$, a sample in $D$ replaces all the ones in $ST$ whose distance from it is less than a given cut-off value $c$. If a sample in $D$ does not replace any sample in $ST$, it is added to $ST$, as it is considered representative of a portion of the features' space that is not covered by pre-existing elements in $ST$. On one side, this implementation speeds up the process of replacement of synthetic samples with real ones; on the other side, depending on the density of the samples in $ST$ and on the cut-off value, it may cause imbalances in the density of samples present in the various regions of the features' space for which $T$ contains information. In fact, a single sample from $D$ may potentially take the place of many others in $ST$.
\vspace*{.2cm}\\~
\textbf{Replace based on Nearest Region (RNR2).} This algorithm represents a variant of RNR. Also RNR2 policy, in fact, only evicts synthetic samples from the training set; however, it differs from RNR in the way samples corresponding to actual measurements are incorporated in the training set. For each element $(x,y)\in ST$, the closest neighbor $(x_r, y_r)\in D$ is found (line 29): if the distance between the two is less than a cut-off value $c$ (line 30), then the output relevant to $x$ is changed from $y$ to $y_r$ (lines 31-32). Like in RNR, if a sample in $D$ does not match any sample in $ST$, it is added to $ST$.
This implementation inherits from RNR the speed in replacing samples in $ST$ with real, new ones, but avoids its downside of changing the density of samples in $ST$: instead of removing samples from $ST$, for each element $(x_r,y_r)$ in $D$, the target value of all the points in the training set for which it is nearest neighbor and within distance $c$ is approximated with $y_r$.

\section{Experimental Evaluation}
\label{sec:expEval}
In this section we  evaluate the various algorithmic and parametric trade-offs discussed in the previous section. To this end we conducts an experimental evaluation based on  two performance critical and widely employed distributed  platforms: a distributed Key-Value Store and a consensus-based coordination service. We start by presenting, in Section~\ref{sec:caseStudy}, the two case studies that will be used throughout the evaluation; then, in Section~\ref{sec:eval:init}, we evaluate our cross-validation-based approach for the construction of the synthetic training set used to bootstrap the gray box model; in Section~\ref{sec:eval:updating} we assess the accuracy achievable by using the different updating algorithms; in Sec.~\ref{sec:eval:sensitivity} we evaluate the robustness of the Bootstrapping technique when the black box model is coupled with AMs delivering different degrees of accuracy; finally, in Sec.~\ref{sec:eval:tuning} we discuss how to identify good values for the tuning parameters of a Bootstrapping-based learner. 

\subsection{Case studies}
\label{sec:caseStudy}

As already mentioned, we consider two case studies:  Infinispan, a popular open-source distributed Key-Value Store (KVS) and a sequencer-based Total Order Broadcast (TOB) service~\cite{Defago2004L}. The choice of these two case studies is motivated by two main reasons. Fist, because of their relevance and wide adoption, they allow to demonstrate the viability of the proposed techniques when applied to mainstream distributed platforms.\remove{First, because of their wide adoption in typical cloud platforms, where they provide performance critical services not only to user applications but also to other fundamental components of the cloud platform.} Second, because of the diversity of the corresponding performance modeling problems: the  features' spaces of the two case studies  have very different dimensionality (2 for TOB vs 7 for KVS), and the corresponding analytical models exhibit different distribution of errors. This allows us to evaluate the proposed solutions in very heterogeneous scenarios, increasing the representativeness of our experimental study.
\subsubsection{Key-Value Store}
\label{sec:kvs}
NoSQL data stores have emerged as popular data platforms for the Cloud. In this study we consider Infinispan, a popular NoSQL open-source data store developed by Red Hat, which, analogously to other recent cloud platforms~\cite{spanner,walterL}, provides a simple, yet highly scalable, key-value data model. In order to enhance performance, Infinispan maintains data fully in-memory and rely on replication as primary mechanism to achieve fault-tolerance and data durability. Finally, similarly to other recent NoSQL cloud data stores~\cite{spanner}, Infinispan provides support for strong consistency via the abstraction of atomic transactions.

Predicting the performance of such platforms is far from being a trivial task, as it is affected by several, often intertwined, factors: contention on physical (i.e., CPU and network) and logical (i.e., data items) resources, characteristics of the transactional workload  (e.g., conflict likelihood and transactional mix) and configuration of the platform itself (e.g., scale and replication degree). This case study is, thus, an example of a modeling/learning problem defined over a large dimensional space (spanning 7 dimensions in our case) and characterized by a complex performance function.

~\\\noindent{\textbf{Base AM.}} The reference model that we employ as base predictor for this case study is PROMPT~\cite{Didona:prompt}. PROMPT relies on the divide-and-conquer approach described in Sec.~\ref{sec:relatedWork}. On one hand, it uses an analytical model that exploits the knowledge of the concurrency and replication scheme (e.g., Two-Phase Commit) employed by the data platform to capture the effects of workload and platform configuration on CPU and data contention via a white box analytical model. On the other hand, it relies on ML to predict latencies of network bound operations. In this study, we pre-train the black-box model used by PROMPT to predict network latencies with a static training set: this means that such model is not updated as samples coming from the running system are collected, thus allowing us to treat PROMPT as a plain white box model.\remove{In this study we pre-train the black-box model used by PROMPT to predict network latencies with a static training set, which allows for achieving very low error ($<10\%$) in all the scenarios included in the test set (to be presented shortly), and which we do not update when we acquire new samples from the operational system. This allows us to treat PROMPT's gray box model analogously to a plain white box analytical one.{\bf Boh, detta cosi' pare che je stai a di' che stamo a bara'. In effetti bariamo x' i campioni di training non li diamo a cubist in realta' ma e' un'altra storia}}
\begin{figure}[t!]
  \centering
\includegraphics[scale=.5]{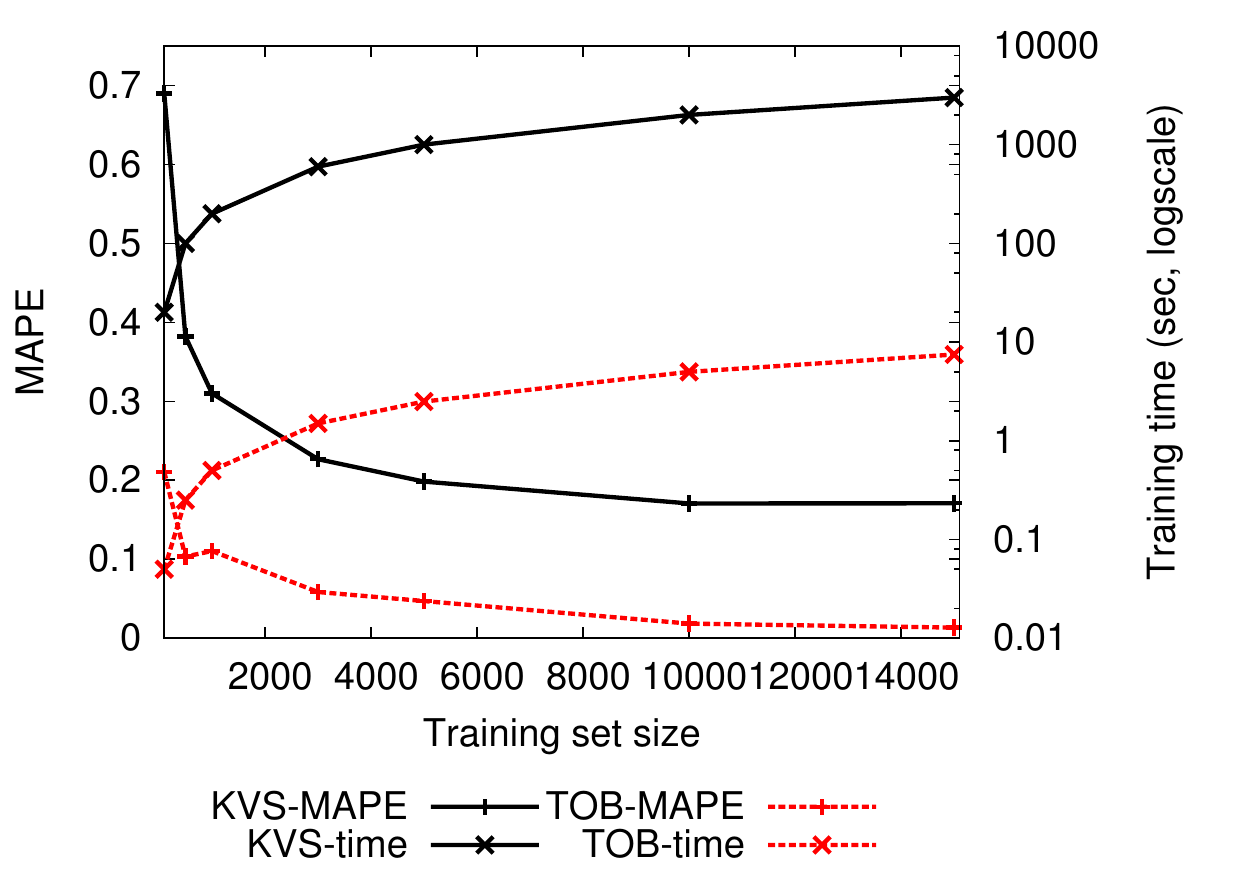}
    \caption{Fitting AM via ML: training time vs MAPE.}
   \label{fig:initialization}
\end{figure}

~\\\noindent{\textbf{Experimental dataset and test bed.}} We consider a dataset composed by approximately nine hundred samples, collected by deploying Infinispan on a private Cloud infrastructure, consisting of 140 VMs deployed over a cluster composed by 18 physical servers equipped with two 2.13 GHz Quad-Core Intel(R) Xeon(R) processors and 32 GB of RAM and interconnected via a private Gigabit Ethernet. The employed virtualization software is Openstack Folsom. The Virtual Machines (VMs) deployed on the cloud are equipped with 1 Virtual CPU and 2GBs of RAM; each VM runs a Fedora 17 Linux distribution with 3.3.4-5.fc17.x86\_64 kernel.

\begin{figure*}[th]
\subfigure[{\bf KVS}: 1K synthetic samples]{\label{fig:mergeKVS1}\includegraphics[width=4.55cm]{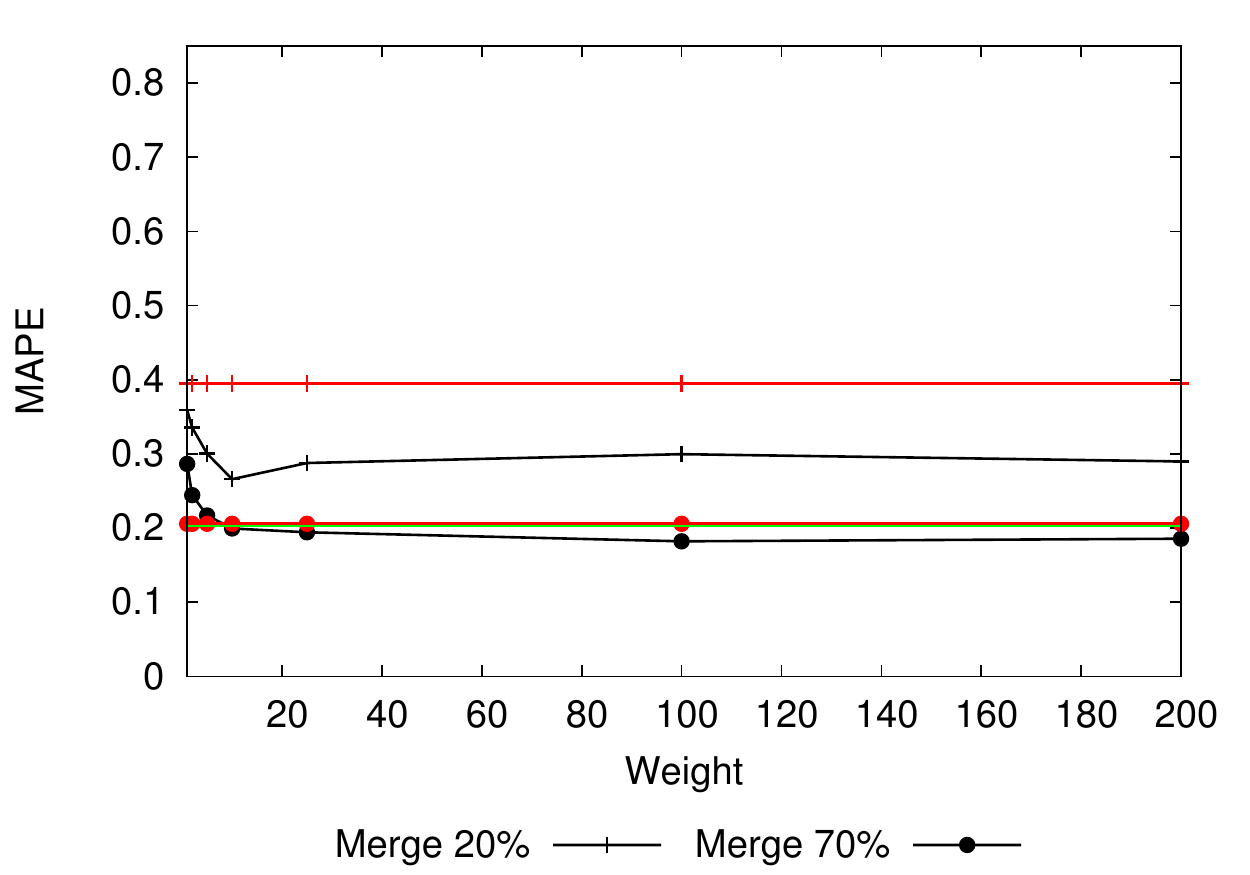}}\hspace{-3.5mm}
\subfigure[{\bf KVS}: 10K synthetic samples]{\label{fig:mergeKVS10}\includegraphics[width=4.55cm]{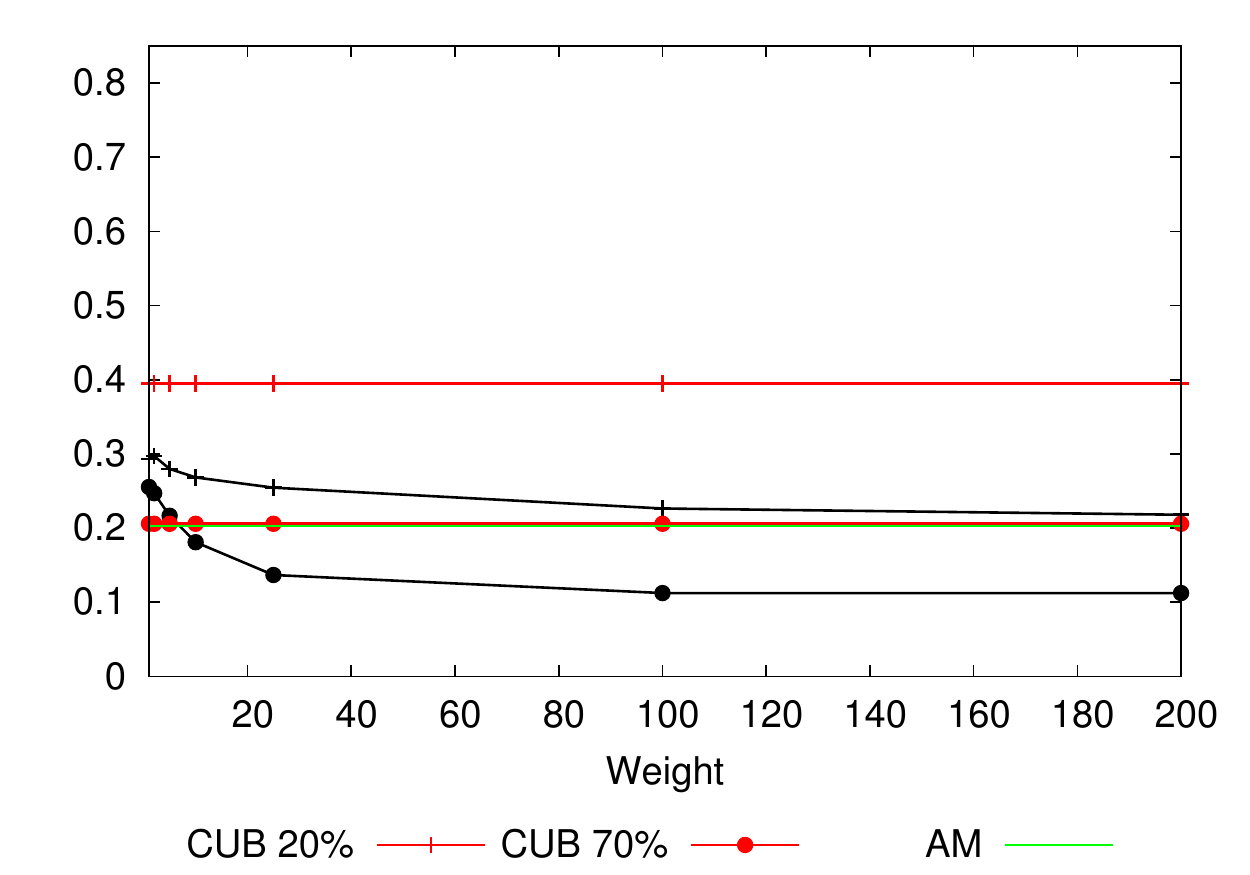}  
}\hspace{-3.5mm}
\subfigure[{\bf TOB}: 1K synthetic~samples]{\label{fig:mergeTOB1}\includegraphics[width=4.55cm]{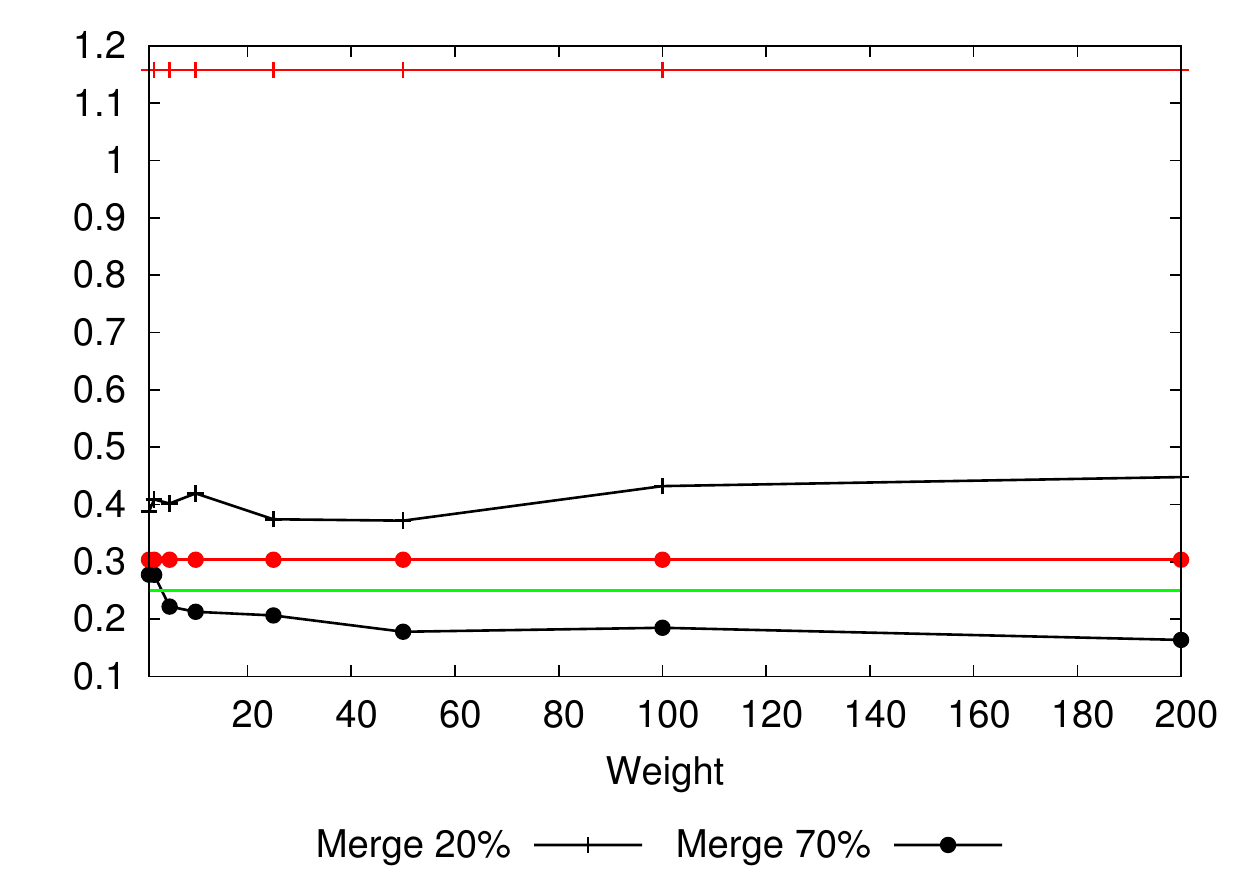}
}\hspace{-3.5mm}
\subfigure[{\bf TOB}: 10K~synthetic samples]{\label{fig:mergeTOB10}\includegraphics[width=4.55cm]{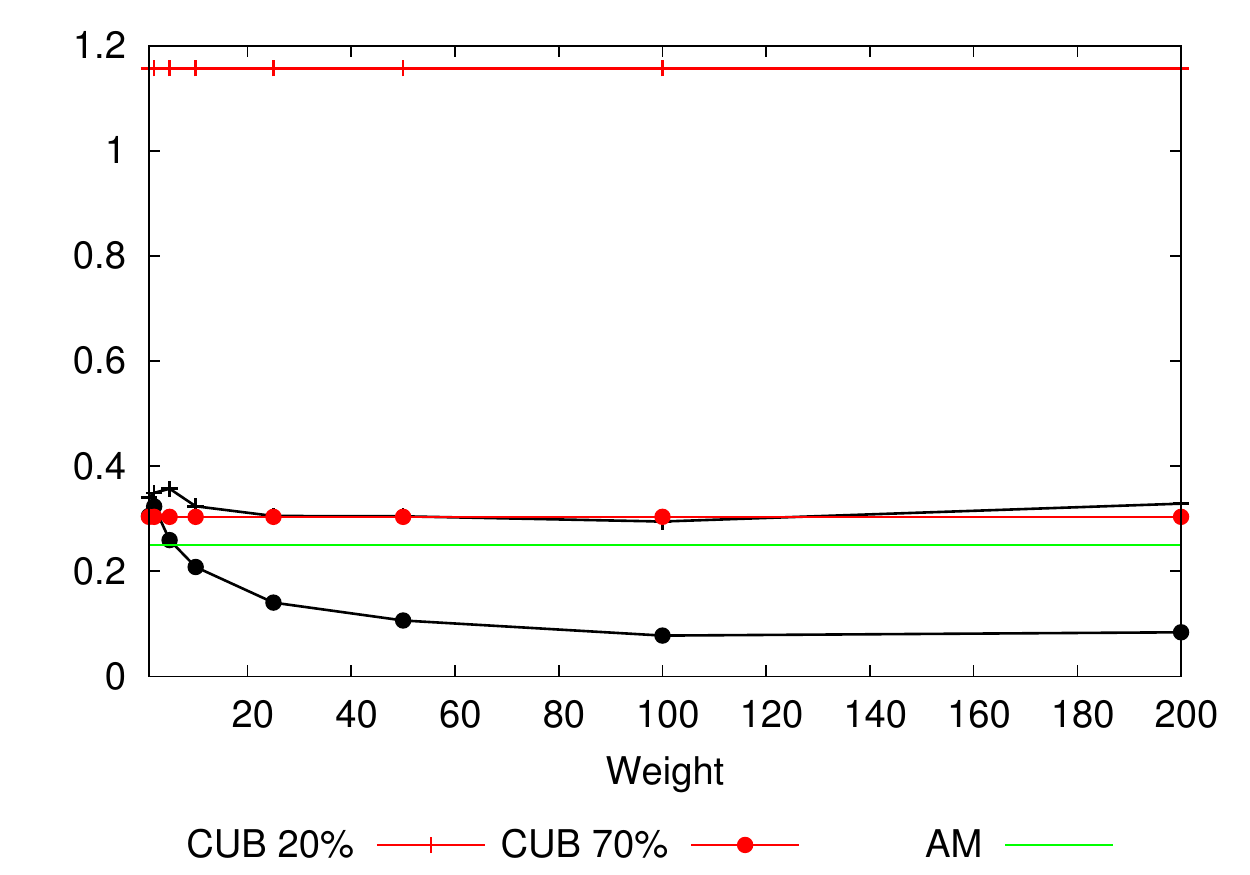}  
}
\caption{Impact of the weight parameter for the Merge updating policy, using 1K and 10K synthetic samples.}
\label{fig:merge}
\end{figure*}

The considered application is a transactional porting of YCSB~\cite{ycsb}, the \emph{de facto} standard benchmark for key-value stores. The dataset consists of YCSB workloads  A, B and F, which were generated using a local thread that injects requests against the collocated Infinispan instance, in closed loop. In order to generate a wider set of workloads, we also let  the number of reads and writes performed by transactions vary between 1 and 5. Finally, we consider two different data access patterns: Zipfian, with zipfian constant 0.7, and Hot Spot, according to which the x\% of the data accesses are biased towards the y\% of the data items (with $x=99$ and $y=1$ in our case); the data set is always composed of 100K keys. 
The samples relevant to the application's throughput are collected while varying workloads and the data platform configuration, deployed on a number of nodes, noted $N$, ranging from 2 to the maximum number of available VMs and set up with a replication factor in the set $\{1,2,3,\frac{N}{2},N\}$.

\subsubsection{Total Order Broadcast}
\label{sec:stob}
Total Order Broadcast is a fundamental building block at the basis of a number of fault-tolerant replication mechanisms~\cite{distributedSystemBookL,PedoneStateMachineL,d2stmL}. We consider a sequencer-based implementation of TOB~\cite{appia}, which generates a message pattern analogous to the one of the Paxos algorithm~\cite{paxos}. Sequencer-based algorithms are probably among the most commonly employed consensus protocols~\cite{appia,distributedSystemBookL,Couceiro10} as they achieve the minimum bound on message latency for these types of problems. On the downside, the sequencer process is typically the bottleneck in these algorithms, as it is required to notify all other nodes in the system of the delivery order of each message disseminated via the TOB primitive. Batching, a.k.a.~message packing~\cite{batchingL}, is a well-known optimization technique that aims at coping precisely with this issue: by buffering messages, the sequencer can amortize the sequencing cost and achieve higher throughput; the message delivery latency however can be negatively affected at low load, due to the additional time spent by the sequencer waiting (uselessly) for the arrival of additional messages. In the following, we denote as $b$ the batching level, i.e., how many messages the sequencer waits to receive before generating a sequencing message. 

~\\\noindent{\textbf{Base AM.} The AM that we adopt as starting point to implement the bootstrapping algorithm is the one described in~\cite{Romano:batchingL}: the sequencer node is abstracted as a $M/M/1$ queue, for which each job corresponds to a batch of messages of size $b$. The message self-delivery latency is computed as the response time for a queue that is subject to an arrival rate $\lambda$ equal to the frequency of arrival of a batch of messages of size $b$ and whose service time $\mu$ accounts both for the CPU time spent for sequencing a message of size $b$ and for the average time waited by a message to see its own batch completed.

~\\\noindent{\textbf{Experimental dataset and test bed.}} We consider a data set containing a total of five hundred observations, corresponding to a uniform sampling of the aforementioned bi-dimensional space, and drawn from a cluster of 10 machines equipped with two Intel Quad-Core XEON at 2.0 GHz, 8 GB of RAM, running Linux 2.6.32-26 server and interconnected via a private Gigabit Ethernet. In the experiment performed to collect the samples, the batching level was varied between 1 and 24, and 512 bytes messages were injected at arrival rates ranging from 1 msgs/sec to 13K msgs/sec.

\subsection{Initialization}
\label{sec:eval:init}
We start our study by evaluating the impact on the gray model's accuracy and construction time depending on the number of samples of the features' space used to populate the initial synthetic training set. We employ, as black box learner, Cubist, a DT regressor that approximates non-linear multivariate functions by means of piece-wise linear approximations~\cite{cubist}. As already mentioned, the Bootstrapping technique can be implemented with any black box learner; after preliminary experimentation with other ML techniques (ANN and SVM), we have opted for using Cubist, because, at least for the considered case studies, it resulted to be significantly easier to tune and to yield the most accurate predictions. 

\begin{figure*}[t]
   \subfigure[{\bf KVS:} 20\% of {\em real} samples.]{\label{fig:replaceKVS20}\includegraphics[width=4.55cm]{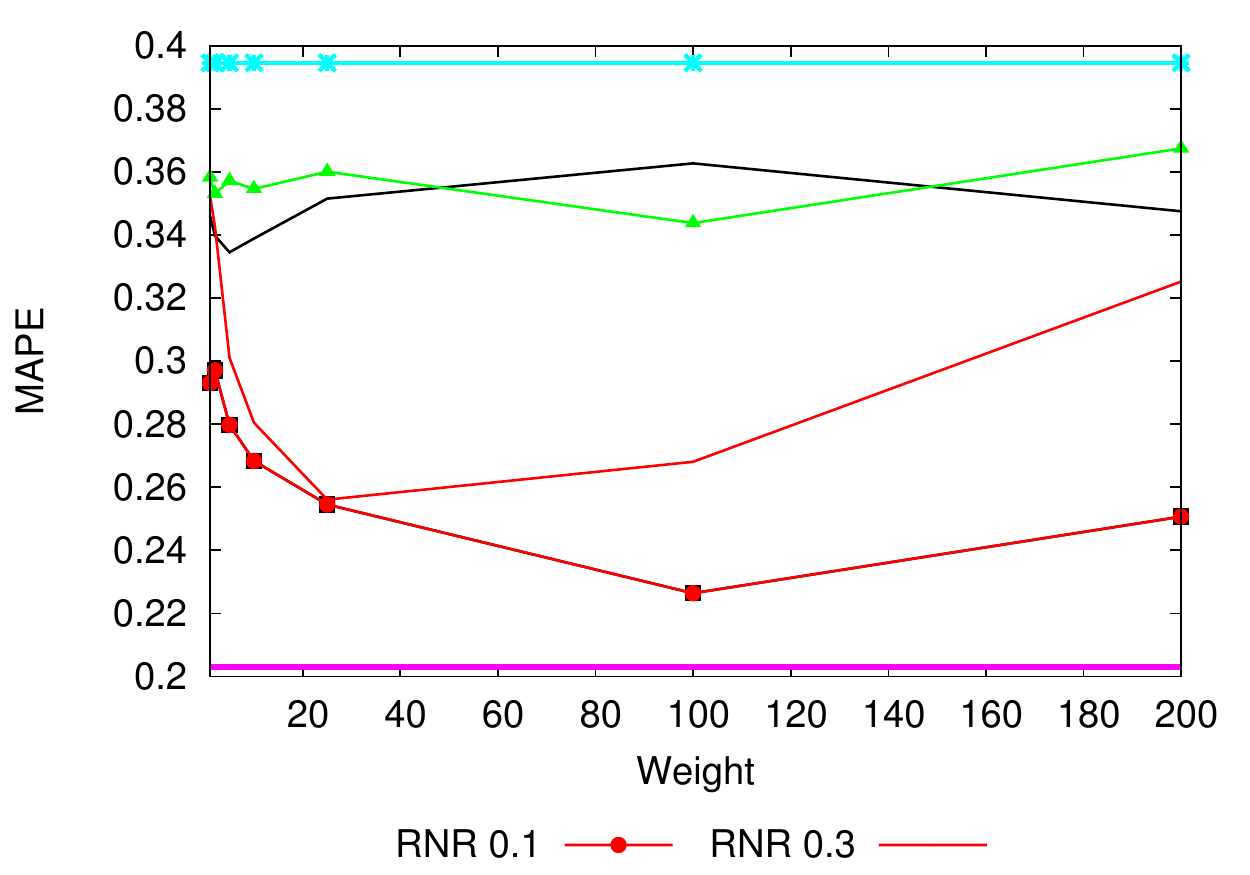}}\hspace{-3mm}
\subfigure[{\bf KVS:} 70\% of {\em real} samples.]{\label{fig:replaceKVS70}\includegraphics[width=4.55cm]{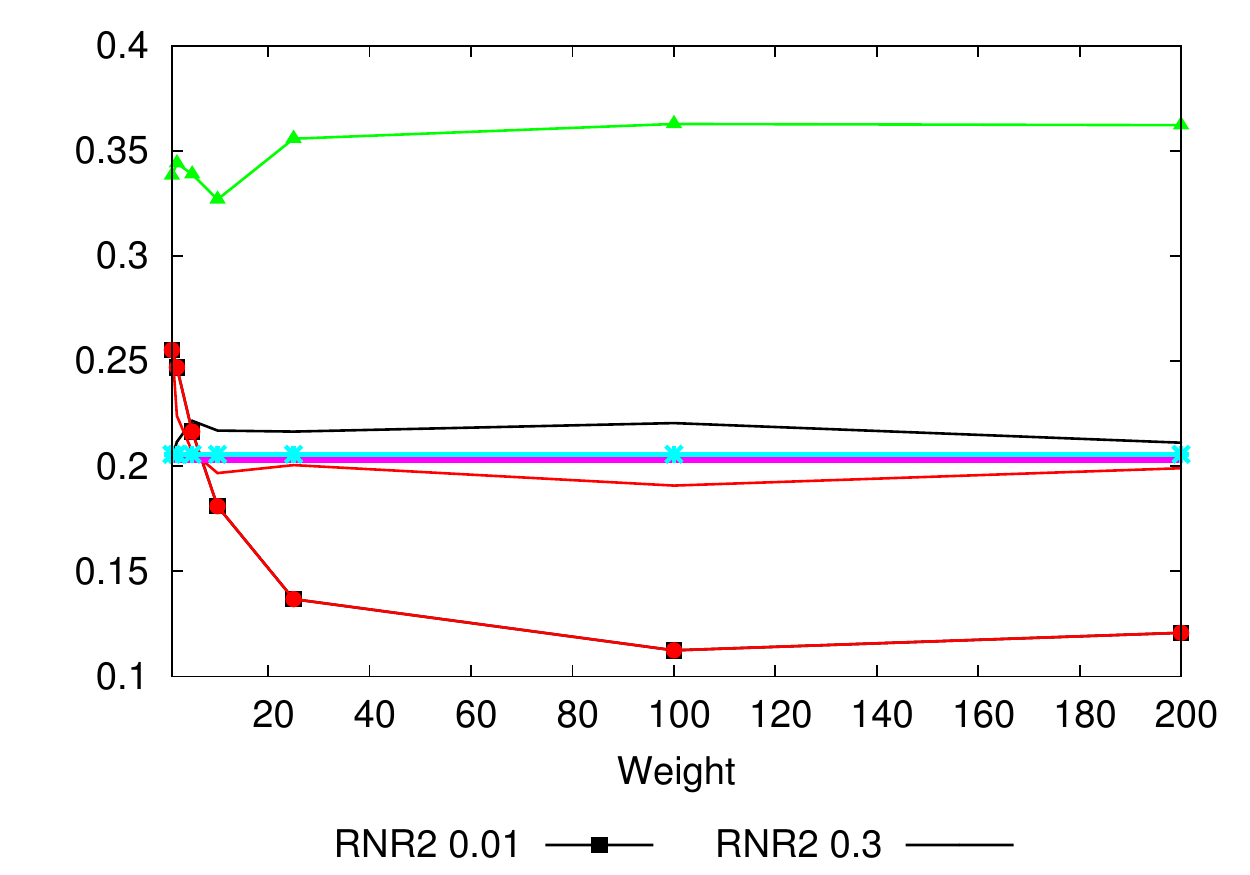}}\hspace{-3mm}
\subfigure[{\bf TOB:} 20\% of {\em real} samples.]{\label{fig:replaceTOB20}\includegraphics[width=4.55cm]{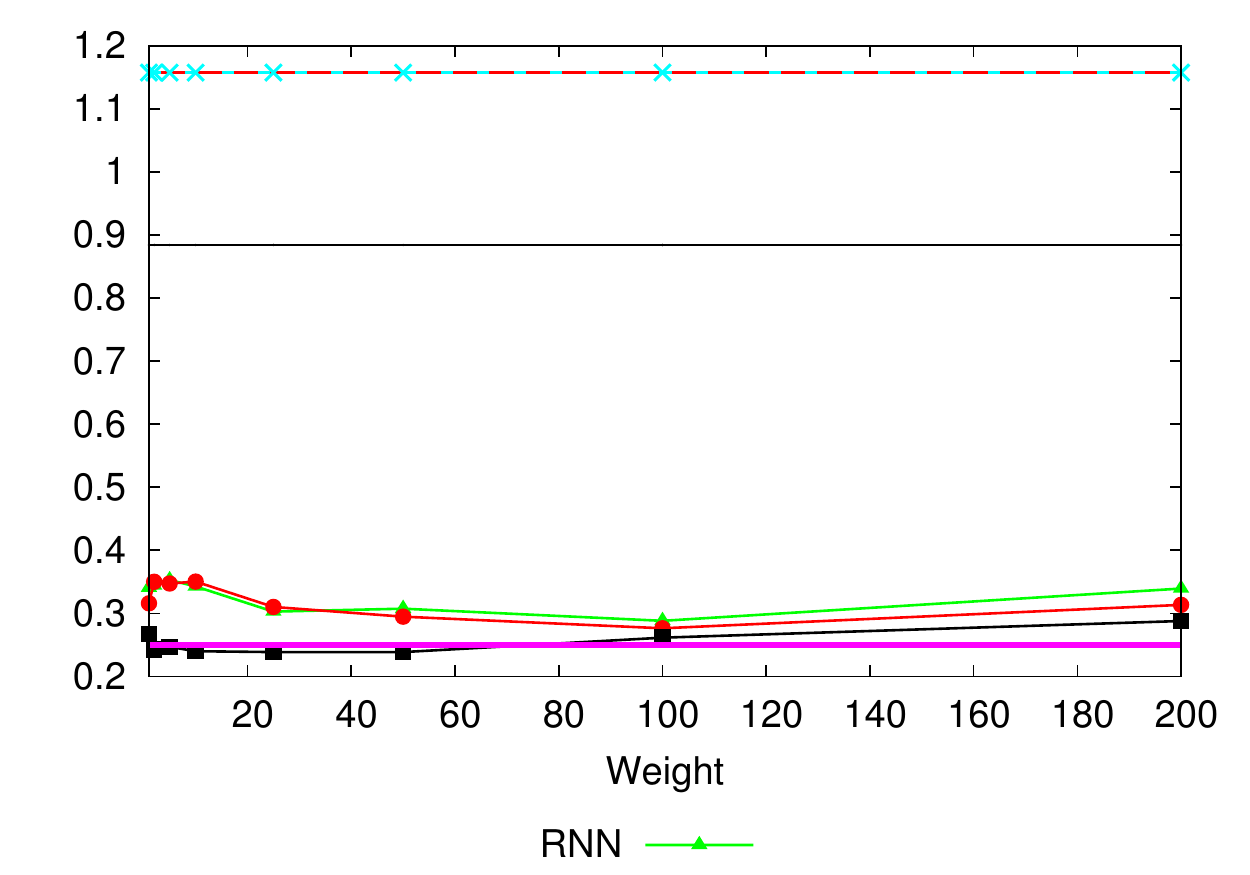}  }\hspace{-3mm}
\subfigure[{\bf TOB:} 70\% of {\em real} samples.]{\label{fig:replaceTOB70}\includegraphics[width=4.55cm]{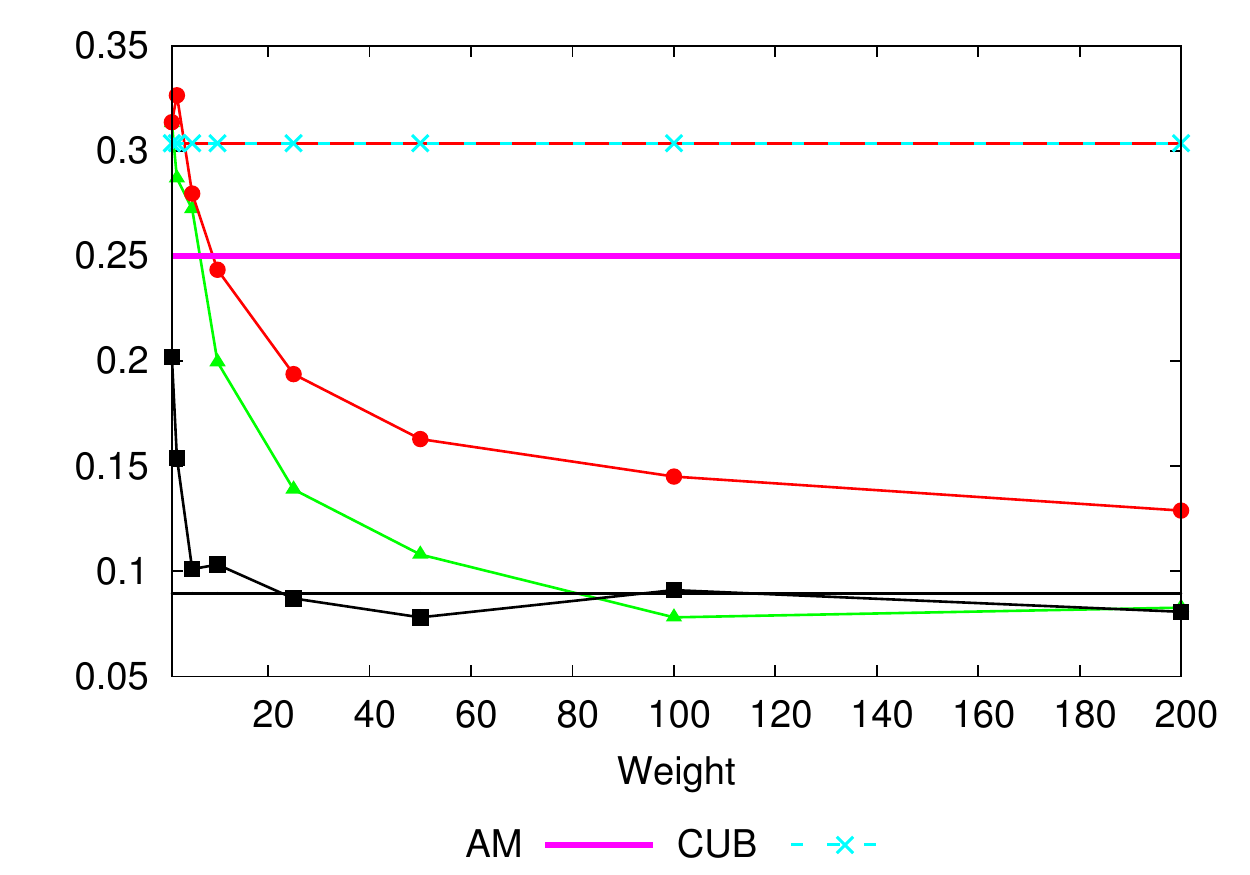}}
\caption{Impact of the weight and cut-off parameters for the RNN, RNR, and RNR2 updating policies, using 10K synthetic samples.}
   \label{fig:replace}
\end{figure*}

Fig.~\ref{fig:initialization} reports, for both case studies, the gray box model building time and the Mean Average Percentage Error (MAPE), computed as~$Avg.(\frac{|real-pred|}{pred})$, of the gray box model with respect to the predictions produced by the AM, evaluated by means of ten-fold cross validation.
On the x-axis we let the number of initial synthetic samples included in the training set of the gray box model vary from 100 to around 15K --- value after which, for both use cases, the ten-fold cross validation accuracy function plateaus. 
The model building time portrayed in the plots corresponds to the sum of the time needed to query the AM in order to generate the synthetic data set of a given cardinality plus the time needed to train the ML over such set. We report that, in our experiments with Cubist, the training time for both case studies has been less than half a second; the gray box model building time in the plots is, thus, largely dominated by the cost needed to query the AM. As shown by Fig.~\ref{fig:initialization}, in the KVS case this cost is much higher than in the TOB one, as the corresponding AM is solved through multiple iterations~\cite{Didona:prompt}. However, it should be noted that the cost to query the AM has to be paid only once, upon initializing the bootstrapped learner, as the update phase only requires to re-train the black-box learner.

Fig.~\ref{fig:initialization} shows that, by fitting the AM using ML techniques, one unavoidably  incurs a loss of accuracy. The actual extent of this accuracy degradation depends on factors such as the number of samples used to construct the initial synthetic training set and the intrinsic capability of the learner to approximate the target function. The plot shows that, as expectable, larger training sets yield a lower approximation error, at the cost of a longer training time; it also shows that Cubist is able to fit the TOB response time function encoded in the analytical model very well (3\% of MAPE with a 10K samples training set) but it is unable to achieve similar accuracy for the KVS case. We argue that this depends on the fact that Cubist approximates non-linear functions by means of piece-wise linear approximation in the leaves of the decision tree that it builds. Such model may be unable to properly approximate the performance function of PROMPT, which is defined over a multi-dimensional space and exhibits strongly non-linear behaviors.  On the other hand, as already mentioned, our preliminary experimentations with alternative learners (ANN and SVM) provided significantly worse approximation errors, especially for the KVS case. This confirms our intuition that the output of PROMPT's AM is indeed a very complex function, which can be hard to approximate using black box learning techniques.

One may argue that the choice of the learner to couple with the AM can be considered another tuning parameter of the Bootstrapping technique. However, identifying the learner that maximizes the prediction accuracy given a training and a test sets is a more general challenge, which falls beyond the sole boundaries of the Bootstrapping technique, and that can be addressed with standard techniques, like Bayesian Optimization~\cite{autowekaL}. Thus, in this paper, we employ Cubist throughout the whole evaluation phase, focusing on the effect of the parameters that are endemic to the Bootstrapping technique. 

Overall, these results highlight that, although ML techniques can typically fit with good accuracy arbitrary functions, they may still introduce approximation errors w.r.t. the original AM. This initial degradation in the accuracy of the gray box model, as we shall see, can actually render it less accurate than the original AM, especially if the gray box model is not fed with a sufficiently large set of additional samples from the operational system.

\subsection{Updating}
\label{sec:eval:updating}
Let us now evaluate the alternative algorithms for the updating of the knowledge base that we presented in Sec.~\ref{sec:update}. We first assess the sensitivity of each algorithm to its key parameters. Finally, we compare their accuracy assuming an optimal parameters tuning.

We start by showing in Fig.~\ref{fig:merge} the results of a study aimed at assessing the impact of the 
weight parameter on the resulting accuracy of the bootstrapped model, while considering synthetic training sets of different initial sizes, namely 1K (Fig.~\ref{fig:mergeKVS1} and~\ref{fig:mergeTOB1}) and 10K samples (Fig.~\ref{fig:mergeKVS10} and~\ref{fig:mergeTOB10}).

We consider two scenarios, in which we assume the availability of 20\% and 70\% of the entire data set composed of collected, real samples, which we feed in input to both the Merge algorithm and to Cubist (non-bootstrapped) that serves as first baseline. As a second reference, we show also the accuracy achieved by using the AM, which incurs a MAPE that is independent of the initial size of the synthetic training set. On the x-axis we vary the weight parameter of the Merge algorithm, and report on the y-axis the MAPE computed with respect the whole set of actual samples (i.e., unlike in the previous section, here the MAPE is not computed with respect to the output of the analytical models).

Concerning the sensitivity to the weight parameter, the plots highlight the relevance of correctly tuning this configuration parameter, especially in the scenario with the larger synthetic training set. In this case, we observe that the best settings of this parameter is relatively larger than for the case of smaller synthetic training set. This can be explained by considering that, by increasing the size of the initial training set, we correspondingly decrease the ratio of real vs synthetic samples (i.e., fabricated by the AM). From the ML perspective this corresponds to decreasing the relevance of the real samples with respect to that of the ``surrounding'' analytical samples. As in this method the analytical samples are never removed from the training set, if the initial synthetic training set is significantly larger than the number of actual samples, these are always surrounded by a large number of synthetic samples, which end up obfuscating the information conveyed by the real ones. By increasing the weight of the samples gathered from the running system, the statistical learner is guided to minimize the fitting error w.r.t. these points. On the other hand, as shown in the case of the small synthetic training set for TOB enriched with 20\% of the set of actual samples (Fig.~\ref{fig:mergeTOB1}) , using excessively large weight values can be detrimental, as it makes the learner more prone to overfitting.

Overall, the experimental data show that  both with large and small initial synthetic training set,  Merge  achieves significantly higher accuracy than both Cubist and the AM, when provided with 70\% of the data in their training set. When the training set percentage is equal to 20\%, the scenario is rather different. In both scenarios, the gray box model still achieves a much higher accuracy than a pure ML-based technique. However, the gray box is only marginally better than the AM with the large initial synthetic training set, and slightly worse than then AM with small initial synthetic training set. This can be explained by considering that the gain achievable using the 20\% training set is relatively small, and can be even outweighed by the loss of accuracy introduced by the learning of the initial AM (see Section~\ref{sec:eval:init}). This is also confirmed by the fact that the MAPE w.r.t. the AM of the gray box model using a synthetic training set of 10K samples is significantly lower than with 1K samples, as shown in Fig.~\ref{fig:initialization}.

\begin{figure*}[th!]
\centerline{
\subfigure[{\bf KVS:} Analytical Model]{\label{fig:h1_kvs}\includegraphics[scale=.4]{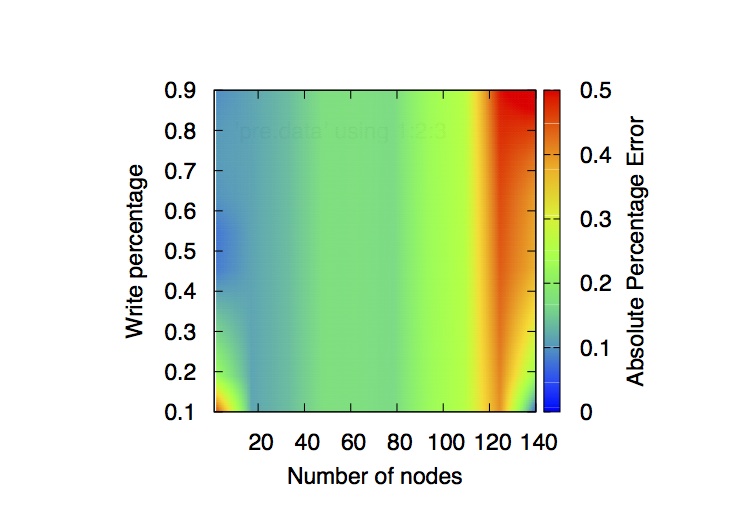}}
\hfil
\subfigure[{\bf KVS:} Cubist]{ \label{fig:h2_kvs}\includegraphics[scale=.4]{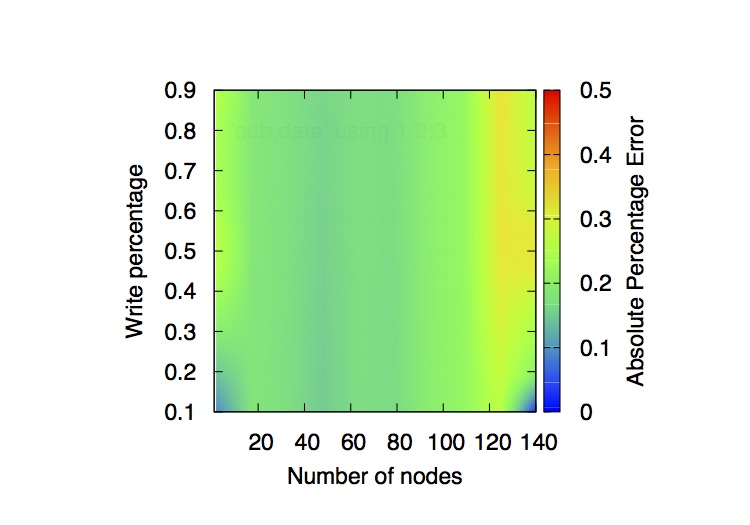}}
\hfil
\subfigure[{\bf KVS:}  Boot. (Merge, weight=100)]{\label{fig:h3_kvs}\includegraphics[scale=.4]{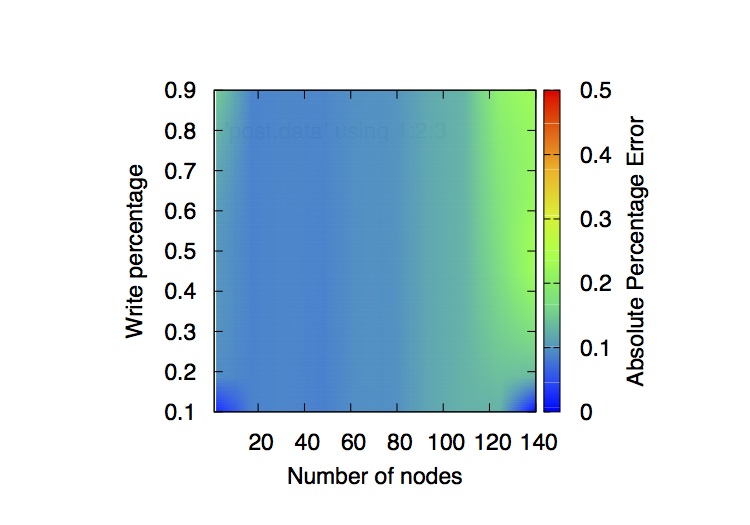}}}
\centerline{
\subfigure[{\bf TOB:} Analytical Model]{\label{fig:h1_tob}\includegraphics[scale=.4]{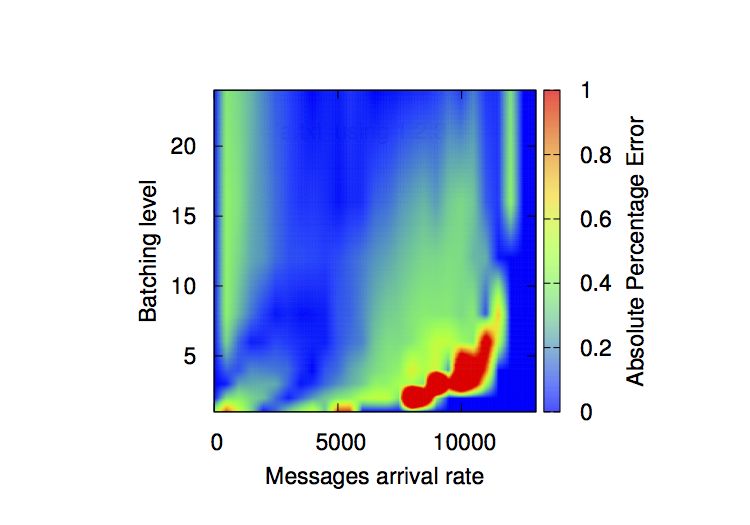}}
\hfil
\subfigure[{\bf TOB:} Cubist]{ \label{fig:h2_tob}\includegraphics[scale=.4]{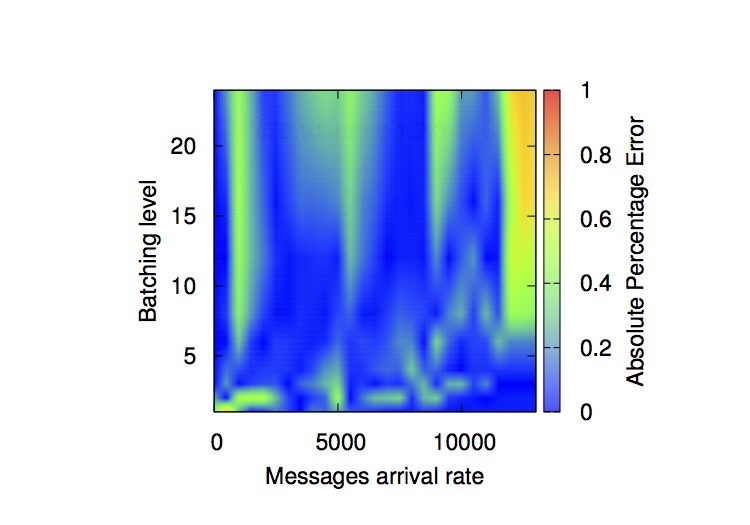}}
\hfil
\subfigure[{\bf TOB:}  Boot. (Merge, weight=100)]{\label{fig:h3_tob}\includegraphics[scale=.4]{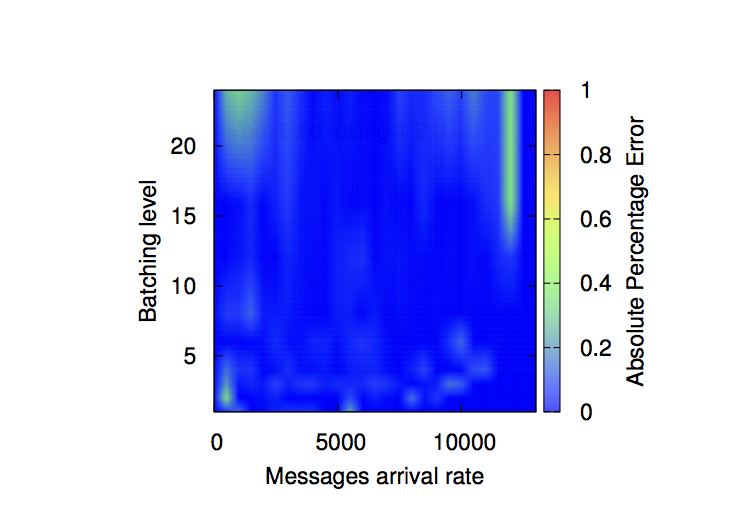}}
}
\caption{Heat map of the Absolute Percentage Error with additional training size 70\%}
\label{fig:heat}
\end{figure*}

In Fig.~\ref{fig:replace} we focus the comparison on the updating policies RNN, RNR, and RNR2. We recall that, unlike Merge, these techniques strive to avoid the coexistence in the training set of ``neighboring'' synthetic and real samples, by removing or replacing synthetic samples close enough to the real samples.
The intuition underlying these approaches is that the information conveyed by the analytical model may be erroneous, and hence contradict the actual samples and confuse the learner. With the exception of the RNN method, which uses exclusively the weight parameter, RNR and RNR2 also use a cut-off parameter, which defines the relative amplitude (normalized w.r.t. a maximum distance) of the radius that is used to determine which synthetic points are to be removed (RNR) or updated  (RNR2), whenever a new real sample is incorporated in the training set. For space constraints, we consider only two values of cut-off, namely 1\% and 30\%, and treat the weight parameter as the independent variable. We choose to report results corresponding to these two cut-off values as, in the light of our experimentation, they are the ones that allow us to best show the impact that this parameter has on a bootstrapped model in the two considered case studies.

Fig.~\ref{fig:replaceKVS20} and ~\ref{fig:replaceTOB20}, resp. Fig.~\ref{fig:replaceKVS70} and ~\ref{fig:replaceTOB70}, report the MAPE achieved when using 20\%, resp. 70\%, of the real data set as training set, reporting, as before, the reference values achieved by the AM and by Cubist (non-bootstrapped). 
The first result highlighted by these plots is that,  also in the replace-based update variants, the weight parameter plays a role of paramount importance. Also the cut-off parameter has a huge impact on the final accuracy of the hybrid model, when implementing RNR and RNR2. 
Moreover we see that the bootstrapped model's accuracy function differs, even fixing the internal parameters, depending on the use case. 

This happens for two main reasons: $i)$ the performance functions output by the AMs for the two use cases exhibit very different trends and are defined over spaces of different dimensionality; $ii)$ the distribution of the real samples w.r.t. the synthetic ones is not the same for the two use cases. For the TOB case, in fact, both real and synthetic samples are drawn uniformly at random from the whole space of possible arrival rate and batching level configurations. Conversely, for the KVS case, the samples in the synthetic training set are drawn uniformly at random but the real ones are not as they are, instead, representative of typical configurations and workloads for that kind of platforms. For example, the density of the points characterized by a number of nodes smaller than 25 is higher than the one relevant to points corresponding to more than 100 nodes in the platform; in the same guise, as already said, the replication degree for data items is defined over the set $\{1,2,3,\frac{N}{2},N\}$, being $N$ the number of nodes. Such asymmetry gives us the possibility to assess the robustness of the Bootstrapping technique 
w.r.t. different densities and distributions of real and synthetic samples.

From the plots we can draw two main conclusions.\vspace*{.1cm}\\
\noindent$i)$ RNN update policy, as described, strives to keep unchanged the initial samples' density in the hybrid training set. Hence, it performs well in the TOB case (Fig~\ref{fig:replaceTOB20} and Fig.~\ref{fig:replaceTOB70}), for which points in the real and synthetic sets are drawn according to the same distribution. On the other hand, it performs poorly in the KVS case, because of the different distribution between real and synthetic samples, and due to the reduced density of synthetic samples in the high dimensional space characterizing the KVS performance function. These factors lead RNN to replace mostly real points in the hybrid set (being them the nearest neighbors of the incoming real samples), instead of evicting synthetic ones. The result, confirmed by the plots in Fig.~\ref{fig:replaceKVS20} and Fig.~\ref{fig:replaceKVS70}, is that the accuracy of the RNN-based bootstrapped model does not increase with the number of real samples gathered from the running system.\vspace{.1cm}\\
\noindent$ii)$ In general, the accuracy of the model bootstrapped with RNR and RNR2 is negatively impacted by excessively large cut-off values, as they are too aggressive in removing knowledge given by synthetic samples. The only exception to this trend is the case for TOB, with 70\% of the real samples and employing the RNR2 updating policy.  Such behavior is clearly shown, for RNR, in Fig.~\ref{fig:replaceTOB20} and Fig.~\ref{fig:replaceTOB70}, corresponding to the TOB use case: with a cut-off value of 0.3, because of the high density of the hybrid training set, RNR evicts all the synthetic samples and replaces them with real ones. The result is that the bootstrapped model delivers the same accuracy as the non-bootstrapped Cubist. This effect is less evident in the KVS case, as the synthetic training set is less dense, and a cut-off of 0.3 is not sufficient to replace all the synthetic samples.
Such behavior is mitigated with RNR2, as this updating policy not only removes synthetic values similar to incoming real ones, but also corrects the output of synthetic samples in the neighborhood. 

Next, in Fig.~\ref{fig:comparison}, we shift perspective, and compare the accuracy achieved by the two best performing updating heuristics, Merge and RNR2, with that achieved by a pure white and black box approach. In this study we set the size of the initial synthetic training set to 10K, and configure the parameters used by Merge using the values that yielded maximum accuracy in the scenarios analyzed so far. This time we change the percentage of real samples observed during the training phase, letting it vary from 10\% to 90\%.

The plot in Fig.~\ref{fig:comparison} clearly highlights the advantages that the bootstrapping technique can provide, outperforming significantly both AM and Cubist, with remarkable gains vs both approaches already at relatively small percentages of training set (30\%-40\%). The data reported in the heat maps in Fig.~\ref{fig:heat} allow us to gain useful insights on the reasons underlying the gains achieved by the Bootstrapping technique vs AM and Cubist. The data in Fig.~\ref{fig:heat} reports the absolute percentage error across the various regions of the features' space achieved by the AM (Fig.~\ref{fig:h1_kvs} and~\ref{fig:h1_tob}), Cubist (Fig.~\ref{fig:h2_kvs} and~ \ref{fig:h2_tob}) and Merge (Fig.~\ref{fig:h3_kvs} and~\ref{fig:h3_tob}). For the case of Merge and Cubist we provide 70\% of the actual data samples as training set, and for Merge we set the weight parameter to 100 and use a synthetic training set of 10K samples. 

For the TOB case, being the corresponding model defined over a two-dimensional space, it is possible to locate exactly the region of the parameters' space where the original AM model incur the highest error. In our case, as depicted in Fig.~\ref{fig:h1_tob}, such region is quite circumscribed in the portion of the heat-map that corresponds to workloads with the highest message arrival rates and low batch value. 
For the KVS case, instead, the error function is defined over the same seven-dimensional features' space of the AM; thus, for visualization purposes, the heat-map corresponds to a projection of the error function over a two-dimensional space defined by the Cartesian product of number of nodes in the platform and percentage of write transactions. Fig.~\ref{fig:h1_kvs} shows that the AM's error is higher in regions corresponding to higher number of nodes and percentage of write transactions.

Fig.~\ref{fig:h3_kvs} and Fig.~\ref{fig:h3_tob} show how, by exploiting both the AM and  the information conveyed by the samples observed from the operational system, the Bootstrapping technique can effectively ``cure'' the errors induced by the AM, and exploit the AM's prediction capabilities in the regions where it performs well, so as to widen the training set for the ML and increase its accuracy.

The AMs employed so far in our study attain a good overall accuracy; with our next experiment, we aim at assessing the impact on the accuracy of a  gray box model bootstrapped with AMs of lower quality. For this experiment, we only consider the TOB case study, as the corresponding analytical model is easier to tamper with in order to reduce its overall accuracy. In fact, the TOB AM relies on the setting of two base parameters, which encapsulate the CPU demands corresponding to the fixed cost of processing a batch of messages and to the cost of processing any additional message in the batch~\cite{Romano:batchingL}. 
As these parameters should be obtained by performing a set of preliminary performance tests, altering their value corresponds to simulating scenarios in which the AM is instantiated with sub-optimally configured parameters as a consequence of noisy or erroneous measurements.

In Fig.~\ref{fig:worse} we treat again the percentage of real samples in the training set as the independent parameter of this study, and consider two models of degraded quality, which achieve, respectively, a MAPE of 35\% (Fig.~\ref{fig:rep1}) and 70\% (Fig.~\ref{fig:rep2}). Also in this case, we consider the Merge and RNR2 variants of the bootstrapping technique, adopting the same parameters used to produce the plot in Fig.~\ref{fig:comparison} and an initial synthetic training set of cardinality 10K. Our experimental data confirm that the gains with respect to a conventional black box learner, such as Cubist, tend to become smaller if the quality of the AM used to bootstrap the learner's knowledge base is weaker. However, and somewhat surprisingly, the Bootstrapping technique can still extract some useful information, and outperform a pure black box approach, even when using very weak analytical models such as the one considered in the right plot of Fig.~\ref{fig:comparison}.

\subsection{Sensitivity to the quality of the AM}
\label{sec:eval:sensitivity}
\begin{figure*}[t!]
\begin{minipage}{.5\textwidth}
\subfigure[{\bf TOB:} Merge vs Replace]{ \label{fig:comparison_tob}\includegraphics[width=4.2cm]{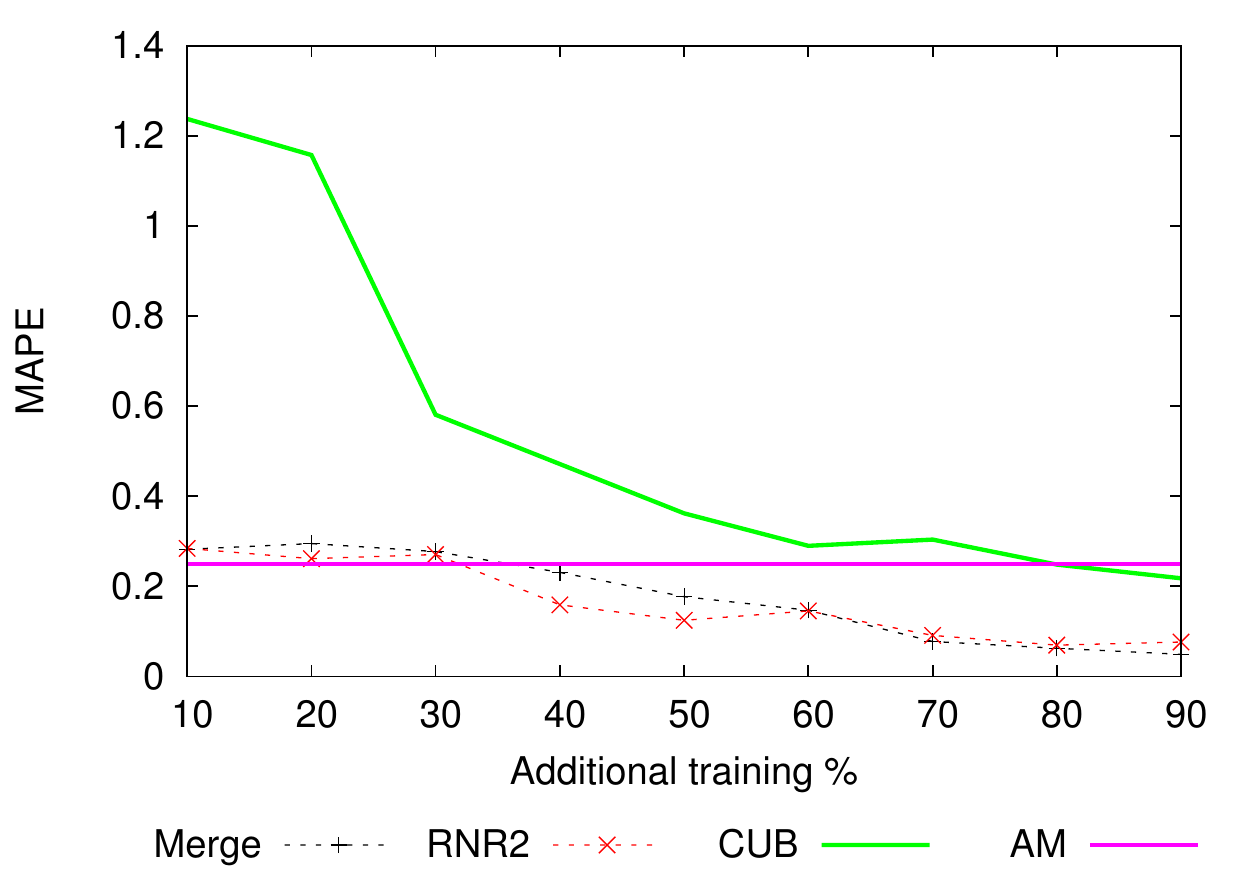}} 
\subfigure[{\bf KVS:} Merge vs Replace]{ \label{fig:comparison_kvs}\includegraphics[width=4.2cm]{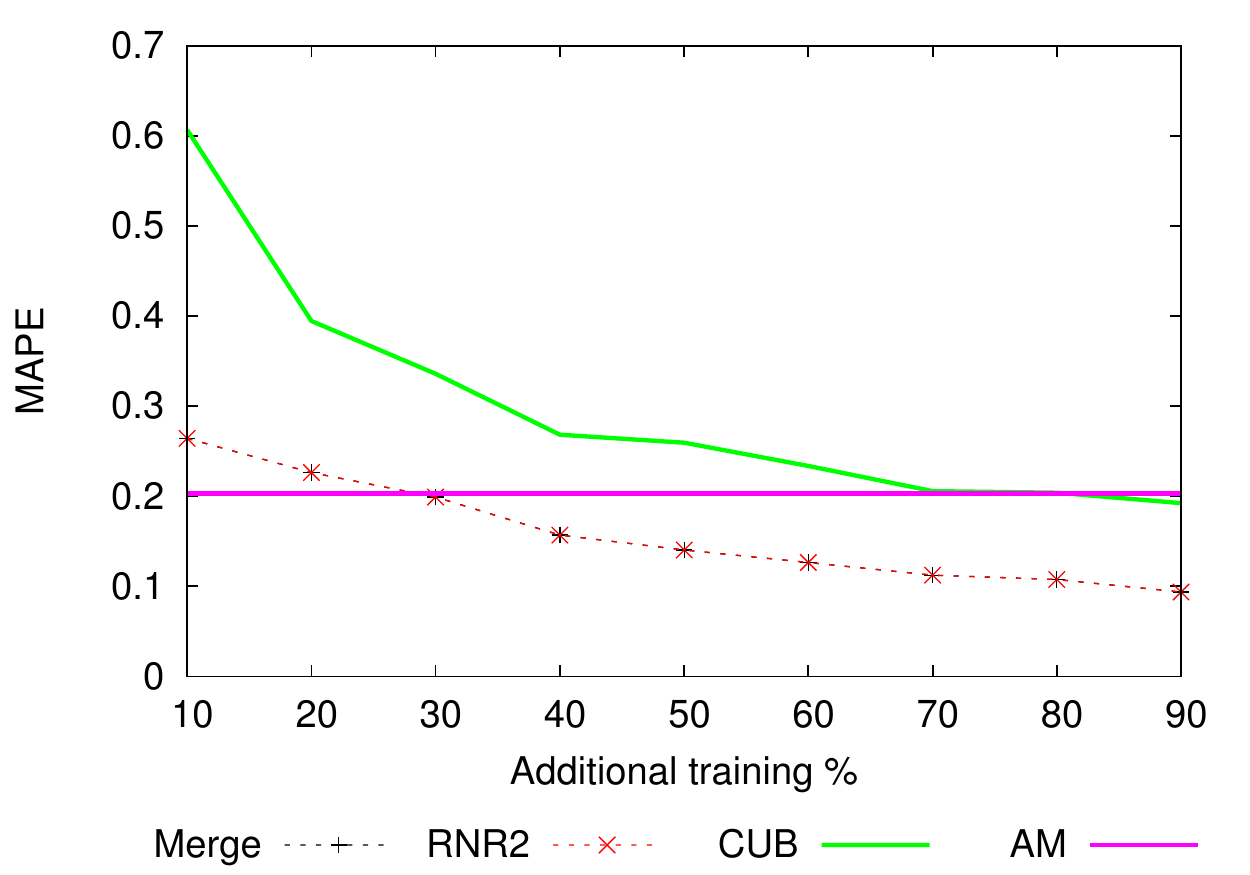}} 
\label{fig:comparison}
\caption{Merge vs Replace (RNR2, cutoff = 1\%), weight = 100}
\end{minipage}
\begin{minipage}{.5\textwidth}
\subfigure[{\bf TOB: }Medium quality AM.]{\label{fig:rep1}\includegraphics[width=4.2cm]{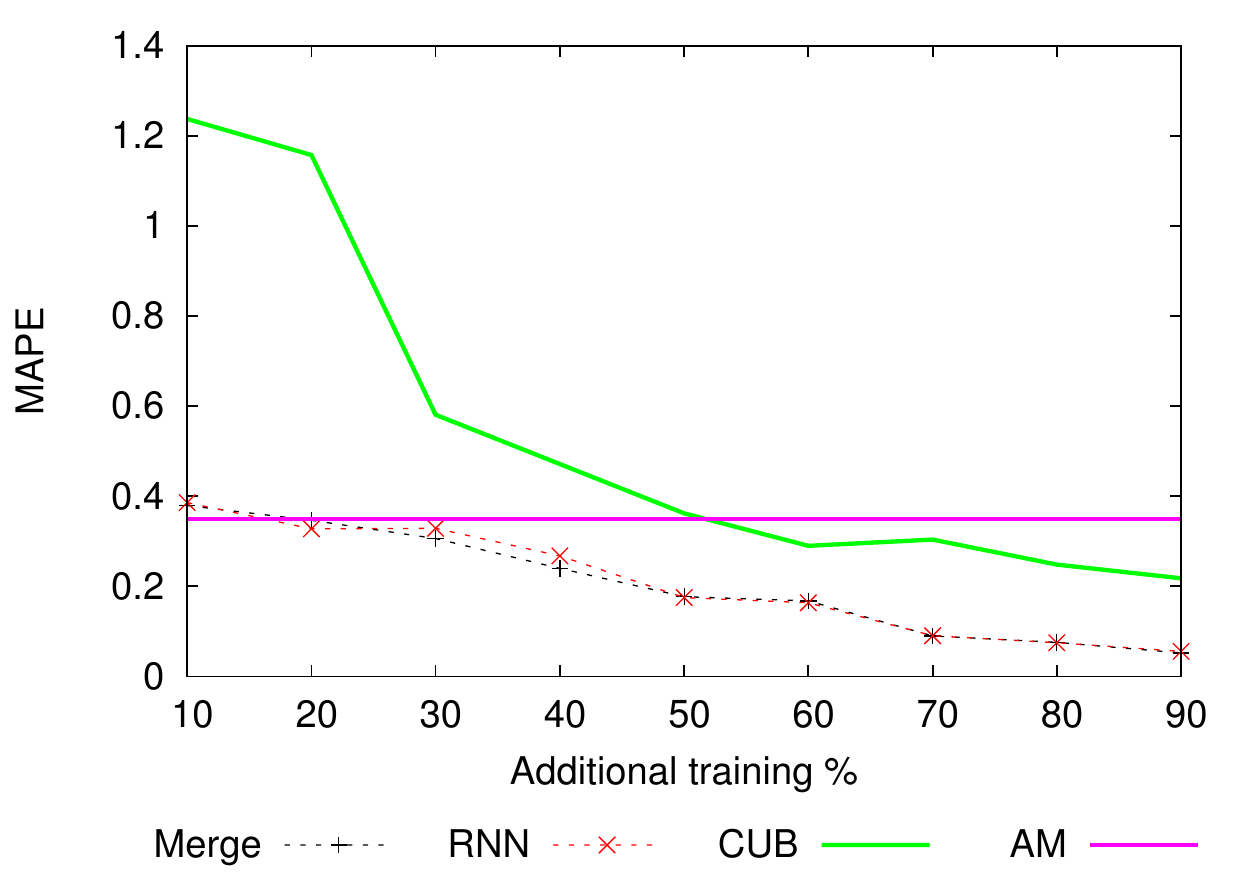}}
\subfigure[{\bf TOB: }Poor quality AM.]{\label{fig:rep2}\includegraphics[width=4.2cm]{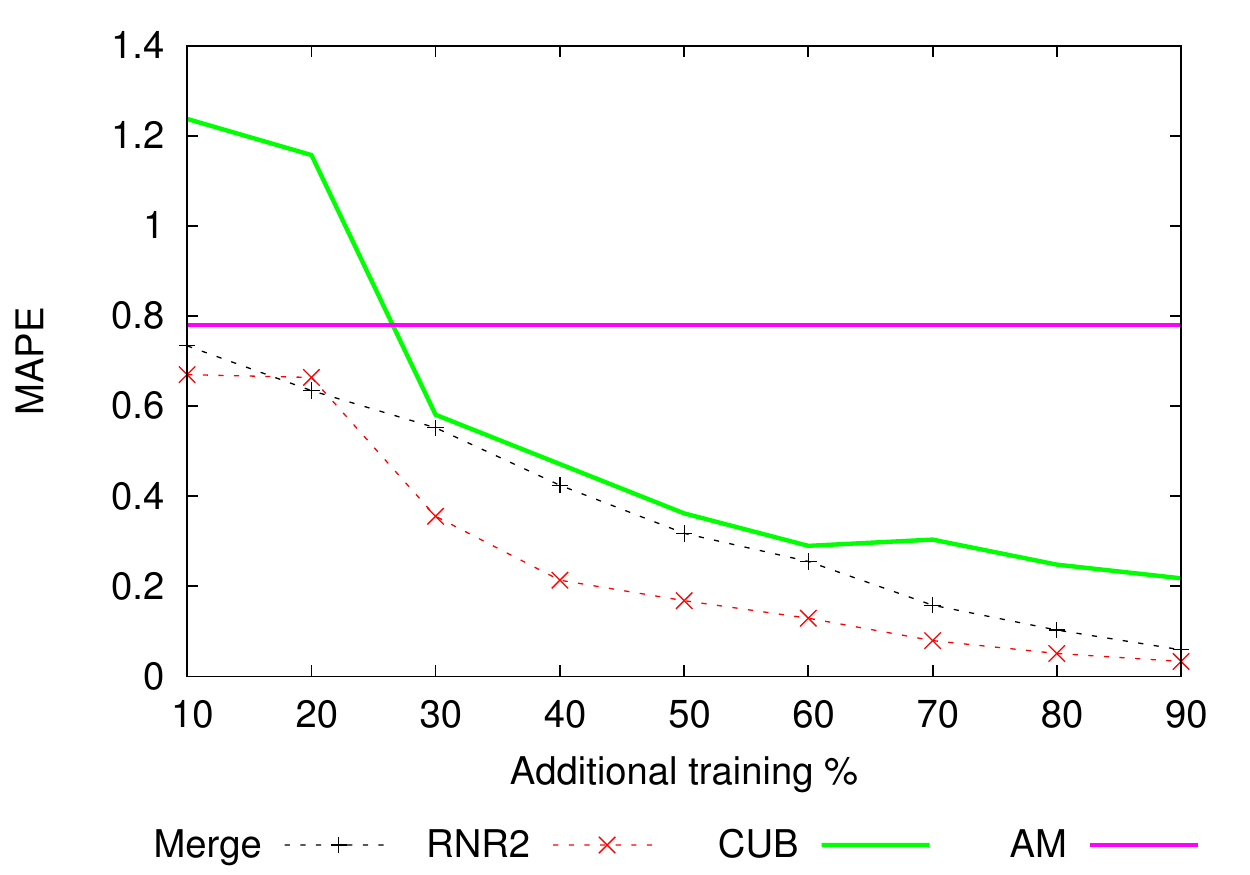}}
\caption{Sensitivity to the usage of AM of different qualities.}
\label{fig:worse}
\end{minipage}
\end{figure*}
Another interesting phenomenon highlighted by the right plot of Fig.~\ref{fig:worse} is the increased gap between the accuracy delivered by RNR2 and Merge, which, so far, had always resulted very close (when optimally tuning the relevant parameters). We argue that this can be explained by considering that RNR2 purges more aggressively than Merge the synthetic samples that fall in proximity of some actual sample. This strategy is clearly the most advantageous in case the employed AM is of mediocre quality.
\subsection{Hyper-parameters optimization}
\label{sec:eval:tuning}
Previous sections have highlighted the sensitivity of the Bootstrapping technique to the setting of its internal parameters: if properly tuned, this technique can  yield considerable gains in terms of accuracy w.r.t. AM and ML employed singularly; conversely, if poorly parametrized, the resulting hybrid model can be worst than the pure black/white box ones at its core.

This is not an idiosyncrasy of the Bootstrapping technique: rather, it is a common characteristic of every black box modeling-based prediction tool. The task of identifying proper values for the internal parameters of a Bootstrapping-based model can be accomplished by employing standard techniques for hyper-parameters optimization proposed in the ML literature, based, for example, on Bayesian optimization or grid/random search~\cite{Bergstra11L}.

\section{Conclusions}
\label{sec:conclusions}
In this paper we have investigated a technique, which we have named Bootstrapping, that aims at reconciling the white box and black box methodologies and at compensating the cons of the one with the pros of the other. 
The design space of the bootstrapping approach includes a number of algorithmic and parametric trade-offs, which can have a strong impact on the accuracy of the resulting gray box model, and which were never identified or discussed in the literature.

In this paper we have filled this gap by presenting what the first detailed algorithmic formalization of this technique. We have identified several crucial choices in the design of Bootstrapping algorithms, proposed a set of alternative approaches to tackling these issues, and evaluated the impact of these alternatives by means of an extensive experimental study targeting two popular distributed platforms (a distributed Key-Value Store and a Total Order Broadcast service).

\bibliographystyle{abbrv}

\end{document}